\documentclass{aa}

\usepackage{amssymb}
\usepackage{amsmath}
\usepackage{txfonts}
\usepackage{graphicx}
\usepackage{xspace}
\usepackage{natbib}
\usepackage{amsmath}
\usepackage{txfonts}
\usepackage{rotating}
\usepackage{url}

\usepackage[usenames]{color}

\newcommand{\her}{\mbox{Her~X-1}\xspace}

\newcommand{\xte}{\textsl{RXTE}\xspace}

\newcommand{\kev}{\ensuremath{\text{ke\kern -0.09em V}}\xspace}
\newcommand{\pca}{\textsl{PCA}\xspace}
\newcommand{\hexte}{\textsl{HEXTE}\xspace}
\newcommand{\asm}{\textsl{ASM}\xspace}

\bibpunct{(}{)}{;}{a}{}{,}

\begin{document}
\title{Probing the outer edge of an accretion disk: A \her turn-on observed
  with \xte}

\author{M. Kuster\inst{1,2} \and J. Wilms\inst{3} \and R.  Staubert\inst{4}
  \and W.A. Heindl\inst{5} \and R. E. Rothschild\inst{5} \and N.I.
  Shakura\inst{6} \and K.A. Postnov\inst{6,7}}

\offprints{M. Kuster, \\ \email{kuster@hll.mpg.de}}

\institute{
Technische Universit\"at Darmstadt, Institut f\"ur Kernphysik,
Schlossgartenstr.~9, 64289 Darmstadt, Germany\and
Max-Planck-Institut f\"ur extraterrestische Physik,
  Giessenbachstr., 85748  Garching, Germany \and 
  Department of Physics, University of Warwick, Coventry, CV7 4AL, U.K.\and 
  Institut f\"ur Astronomie und Astrophysik, Sand 1, 72076 T\"ubingen,
  Germany \and 
  Center for Astrophysics and Space Sciences, UCSD, La Jolla, CA 92093, USA \and
  Sternberg Astronomical Institute, Moscow State University, 119899 Moscow,
  Russia 
  \and Faculty of Physics, Moscow State University, 119899 Moscow, Russia}

\date{Received 12. November 2004 / Accepted 29. June 2005 } 

\titlerunning{Her~X-1 Turn-On}
 
\abstract{We present the analysis of Rossi X-ray Timing Explorer (\xte)
  observations of the turn-on phase of a 35\,day cycle of the X-ray binary
  Her~X-1. During the early phases of the turn-on, the energy spectrum is
  composed of X-rays scattered into the line of sight plus heavily absorbed
  X-rays. The energy spectra in the 3--17\,\kev range can be described by a
  partial covering model, where one of the components is influenced by
  photoelectric absorption and Thomson scattering in cold material plus an
  iron emission line at 6.5\,\kev. In this paper we show the evolution of
  spectral parameters as well as the evolution of the pulse profile during
  the turn-on. We describe this evolution using Monte Carlo simulations
  which self-consistently describe the evolution of the X-ray pulse profile
  and of the energy spectrum.
  
  \keywords{stars: individual (Hercules X-1) -- X-rays: binaries -- stars:
    neutron -- Accretion, accretion disks -- Scattering} }

\maketitle

\section{Introduction}\label{sect:intro}
\object{Her X-1} is one of the best understood X-ray binary systems showing
a variety of long and short term periodicities. The X-ray pulsar spins with
a 1.24\,s period and moves in a 1.7\,day almost circular orbit around its
companion HZ Her \citep[][]{tananbaum:72a}.  Both effects cause a
modulation of the observed flux in optical as well as in the X-rays. In
addition the X-ray light curve of \her shows a long term 35 day intensity
variation. This modulation is the best evidence for an inclined,
precessing, and warped accretion disk in an X-ray binary system. The origin
of the warping is not yet fully understood, but may be caused either by
radiation driven accretion disk winds \citep[][]{schandl:94a} or by
radiation pressure \citep[][]{maloney:97a}. The precessing motion of the
disk can be understood in the context of tidal interaction and/or as a
consequence of non vanishing torques acting on the disk, e.g. due to a
coronal wind \citep[][]{schandl:94a,schandl:97a,shakura:98a,ketsaris:01a}.
Because of the high inclination of the system, the disk periodically blocks
the line of sight to the neutron star during about 60\% of the 35 day
cycle.

The observed 35 day light curve shows two maxima in intensity: the
``main-on'' and the ``short-on'' \citep[][]{giacconi:73a}. Following the
often adopted baseline model of \her \citep[][and references
therein]{katz:73a,schandl:94a,scott:00a,ketsaris:01a,leahy:04b}, the
$\sim$$12\,\text{day}$ long main-on starts when the outer rim of the
accretion disk opens the line of sight to the central neutron star (see
Fig.~\ref{fig:disk-corona-model}).  Subsequently, at the end of the main-on
the inner edge of the accretion disk covers the line of sight.  The second
maximum in intensity occurs as soon as the inner edge of the accretion disk
uncovers the line of sight during the progression of the 35 day cycle. This
phase is called short-on where only $\sim$$35\%$ of the main-on intensity
is measured.  The states in between the short-on and the main-on are called
``low-states'' where the intensity drops to $\sim$$3\%$ of the main-on
intensity \citep[][]{scott:99a}.  The transitions from the low-states to
the main-on and short-on are called ``turn-on'', while the decline in
intensity at the end of the main-on and the short-on generally is called
``turn-off''. This periodic behavior is irregularly interrupted by
anomalous low states as it was observed in 1983 \citep[][]{parmar:85a},
1993 \citep[][]{vrtilek:94a,vrtilek:96a}, 1999
\citep[][]{parmar:99a,coburn:00a,oosterbroek:01a}, and recently in 2004
\citep[][]{boyd:04a}. During an anomalous low state the maximum X-ray flux
typically drops to 1--3\% of the main-on flux.

While the end of the main-on has been the subject of previous
\textsl{Ginga} observations \citep[e.g.,][]{deeter:98a}, observational data
on the spectral evolution during the start of the main-on has been rare.
Observations of turn-ons of \her have been presented by \citet{becker:77a},
\citet{davison:77a}, and \citet{parmar:80a} with different instruments. All
of these observations show a strong indication that the flux during the
early stages of the turn-on is composed of heavily absorbed plus scatted
radiation.
\begin{figure*}
  \centering
  \includegraphics[width=17cm]{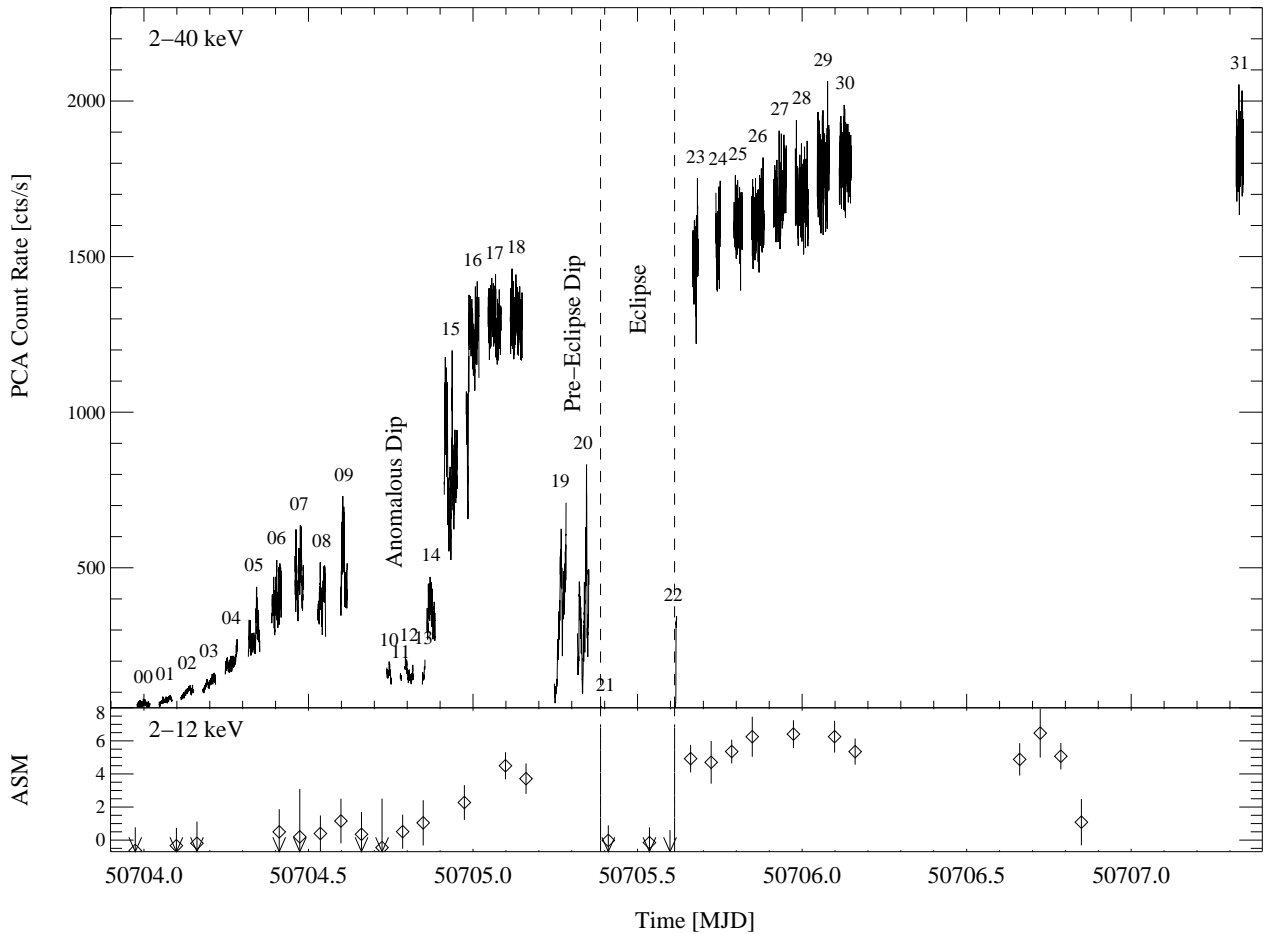}
  \caption{From top to bottom: \xte \pca and  \xte \asm count rates for the
    time of the turn-on. The beginning and the end of an eclipse are marked
    by dashed lines. The numbers identify individual \xte orbits starting
    with orbit~00 to orbit~31.}\label{turnonlc}
\end{figure*}
In 1997 September we observed a complete turn-on of the 35 day cycle of
\her in a two day continuous observation with the Rossi X-ray Timing
Explorer (\xte). In this paper we present results from the spectral and
temporal analysis of this observation. Further we describe these data with
a physical model that can reproduce the spectral features and temporal
evolution of the pulse shape seen in this observation. For our physical
model we assume that the variations of the observed spectrum are due to the
varying column density caused by cool gas of the outer rim of the accretion
disk and due to radiation scattered from (ionized) gas sandwiching the
accretion disk (an accretion disk corona).

The remainder of this paper is structured as follows: In
Section~\ref{sec:datareduction} we give a short description of our
observation and the extraction of the data, before we describe our spectral
model and the results of the spectral analysis in
Section~\ref{sec:specevol}, we present the results of our analysis of the
evolution of the pulse profile in Section~\ref{sec:pulsevol}. In
Section~\ref{sec:pulsesim} we introduce a method to determine the amount of
absorbed and scattered radiation by simulating the influence of a
scattering medium on the shape of the pulse profile using Monte Carlo
simulations. We summarize this paper in Section~\ref{sec:conclusion} and
propose a model of the outer accretion disk rim which can explain the
spectral behavior as well as the evolution of the pulse profile.
\begin{table}[tH]
\caption{Observing log of the Turn-On observations of Her~X-1.\label{tab:x3log}}
\begin{center}
\begin{tabular}{crcr}\hline\hline
 Obs. &\multicolumn{1}{c}{Date} & Exposure &\multicolumn{1}{c}{Count Rate} \\ 
 & \multicolumn{1}{c}{[MJD]}   & [sec]  & \multicolumn{1}{c}{[counts\,s$^{-1}]$} \\
\hline
\noalign{\vskip 2pt}
00 & 50703.979 &
~~ 3300 &  $ ~ 62.1\pm0.2$ \\ 
\noalign{\vskip 2pt}
01 & 50703.979 & 
~~ 3400 &  $ ~ 73.3\pm0.3$ \\ 
\noalign{\vskip 2pt}
02 & 50703.979 & 
~~ 3300 &  $ 100.0\pm0.3$ \\ 
\noalign{\vskip 2pt}
03 & 50703.979 & 
~~ 3300 &  $ 130.0\pm0.3$ \\ 
\noalign{\vskip 2pt}
04 & 50703.979 & 
~~ 3200 &  $ 195.4\pm0.3$ \\ 
\noalign{\vskip 2pt}
05 & 50704.312 & 
~~ 3000 &  $ 280.8\pm0.4$ \\ 
\noalign{\vskip 2pt}
06 & 50704.312 & 
~~ 2600 &  $ 402.7\pm0.5$ \\ 
\noalign{\vskip 2pt}
07 & 50704.452 & 
~~ 2300 &  $ 469.2\pm0.5$ \\ 
\noalign{\vskip 2pt}
08 & 50704.452 & 
~~ 2100 &  $ 395.0\pm0.5$ \\ 
\noalign{\vskip 2pt}
09 & 50704.452 & 
~~ 1700 &  $ 501.7\pm0.6$ \\ 
\noalign{\vskip 2pt}
10 & 50704.727 & 
~~ 1200 &  $ 158.8\pm0.5$ \\ 
\noalign{\vskip 2pt}
11 & 50704.727 & 
~~  300 &  $ 148.7\pm1.0$ \\ 
\noalign{\vskip 2pt}
12 & 50704.727 & 
~~ 2100 &  $ 160.6\pm0.4$ \\ 
\noalign{\vskip 2pt}
13 & 50704.727 & 
~~  700 &  $ 156.7\pm0.7$ \\ 
\noalign{\vskip 2pt}
14 & 50704.860 & 
~~ 2100 &  $ 358.4\pm0.5$ \\ 
\noalign{\vskip 2pt}
15 & 50704.860 & 
~~ 3400 &  $ 813.6\pm0.5$ \\ 
\noalign{\vskip 2pt}
16 & 50704.860 & 
~~ 3300 &  $1197.0\pm0.6$ \\ 
\noalign{\vskip 2pt}
17 & 50705.045 & 
~~ 3400 &  $1298.1\pm0.7$ \\ 
\noalign{\vskip 2pt}
18 & 50705.045 & 
~~ 3200 &  $1293.7\pm0.7$ \\ 
\noalign{\vskip 2pt}
19 & 50705.045 & 
~~ 3100 &  $ 325.6\pm0.4$ \\ 
\noalign{\vskip 2pt}
20 & 50705.312 & 
~~ 3000 &  $ 329.3\pm0.4$ \\ 
\noalign{\vskip 2pt}
21 & 50705.381 & 
~~ 2700 &  $ ~ 34.9\pm0.3$ \\ 
\noalign{\vskip 2pt}
22 & 50705.591 & 
~~ 1800 &  $ ~ 73.2\pm0.4$ \\ 
\noalign{\vskip 2pt}
23 & 50705.659 & 
~~ 1500 &  $1494.1\pm1.1$ \\ 
\noalign{\vskip 2pt}
24 & 50705.726 & 
~~ 1200 &  $1568.2\pm1.2$ \\ 
\noalign{\vskip 2pt}
25 & 50705.793 & 
~~ 2100 &  $1607.3\pm0.9$ \\ 
\noalign{\vskip 2pt}
26 & 50705.793 & 
~~ 3300 &  $1612.5\pm0.7$ \\ 
\noalign{\vskip 2pt}
27 & 50705.793 & 
~~ 3400 &  $1697.5\pm0.7$ \\ 
\noalign{\vskip 2pt}
28 & 50705.793 & 
~~ 3400 &  $1699.1\pm0.7$ \\ 
\noalign{\vskip 2pt}
29 & 50705.793 & 
~~ 3100 &  $1787.7\pm0.8$ \\ 
\noalign{\vskip 2pt}
30 & 50706.096 & 
~~ 3100 &  $1810.4\pm0.8$ \\ 
\noalign{\vskip 2pt}
31 & 50707.313 & 
~~ 1900 &  $1828.2\pm1.0$ \\ 
\noalign{\vskip 2pt}
\hline\end{tabular}\end{center}
{\small Exposure times shown are rounded to the closest
100\,sec. The count rate is background subtracted.}
\end{table}

%%% Local Variables: 
%%% mode: latex
%%% TeX-master: "~/publ/diss/diss"
%%% End: 

\section{Observation and data reduction}\label{sec:datareduction}
We observed \her for two days on 1997 September 13/14 with \xte.
Fig.~\ref{turnonlc} shows the \xte Proportional Counter Array (\pca) light
curve of the entire observation, together with the light curve measured
simultaneously by the \xte All Sky Monitor (ASM). Note that during the
entire observation all five PCUs of the \xte PCA were active and therefore
throughout this paper the count rates are consistently given in
$\text{cts}/\text{s}$ for five PCUs. During the turn-on an eclipse took
place around MJD~50705.5. Furthermore, two dips were detected: a
pre-eclipse dip around MJD~50705.3 and an anomalos dip around MJD~50704.8.
The gaps in the light curve are due to Earth occultations and SAA passages
during individual \xte orbits. The exposure times, mean count rates, and
dates of observation are given in Table~\ref{tab:x3log} for each \xte
orbit.

For the detailed analysis, we extracted \pca light curves for all \xte
orbits listed in Table~\ref{tab:x3log} with a time resolution of 16 ms.
Light curves were extracted for five energy bands: 2.0--4.5\,\kev,
4.5--6.5\,\kev, 6.5--9\,\kev, 9--13\,\kev, and 13--19\,\kev. After
correcting the photon arrival times with respect to the solar systems
barycenter and for the orbital motion of the neutron star, we determined
the pulse period of \her by folding the data of the 13--19\,\kev energy
band using a $\chi^2$ maximization test.  The resulting pulse period is
$P_{\text{Spin}}=1.2377291(2)\,\text{s}$ (MJD~50708.199), which is
consistent with observations of, e.g., \cite{dalfiume:98a} and
\cite{coburn:00a}.  Subsequently we folded all light curves with this pulse
period to obtain a pulse profile for each energy band and \xte orbit. Pulse
phase $\Phi_{1.24}=0$ was defined as the time of the maximum flux in the
main pulse of the profile in the energy range 13--19~\kev. From each
profile we subtracted the unpulsed flux and normalized the count rate to
the maximum.  Fig.~\ref{fig:pulspanel} shows the evolution of the pulse
profiles in different energy ranges over the time of the whole turn-on. We
have omitted those orbits during which the pulse profile shows no
remarkable variation compared to the previous or following orbit, i.e., the
orbits~01, 06--08, 16--18, and 25--30.

For the spectral analysis, pulse phase averaged \pca and \hexte spectra
were extracted for each individual \xte orbit. To minimize the background,
we have chosen good-time intervals (GTI) with an ``electron-ratio'' of all
PCUs less than 0.1 \citep[see e.g.,][]{wilms:99a}. All spectra are
background and dead-time corrected.  The data of orbits 10--14 and 19--22
were omitted for the analysis because our spectral model is not applicable
during the times of the dips and the eclipse. We simultaneously fitted our
spectral model described in section~\ref{subsec:model} to both the \hexte
and \pca data of each orbit. The systematic uncertainties of the response
matrix of the \pca assumed for the spectral analysis are given in
Table~\ref{tab:pca-systematics}.
\begin{table}
  \caption{Systematic errors applied to the \pca data to account for
    uncertainties in the \pca response matrix.}\label{tab:pca-systematics}
\centering
    \begin{tabular}{rrc}\hline\hline
      Channel               & Energy Range & Systematics \\ \hline
      0--15                 & 2--8\,\kev    & 1.0\%  \\
      16--39                & 8--18\,\kev   & 0.5\%  \\
      40--57                & 18--29\,\kev  & 2.0\%  \\
      58--128               & 29--120\,\kev & 5.0\%  \\\hline
    \end{tabular}
\end{table}

\section{Evolution of spectral parameters}\label{sec:specevol}
Before we present the results of our spectral fitting in
Sect.~\ref{subsec:specevol}, we give an introduction to the complex
spectral model used to describe the data.

\subsection{Spectral model}\label{subsec:model}
As earlier observations already have shown
\citep[][]{davison:77a,parmar:80a,becker:77a}, a combination of direct,
scattered, and absorbed photons is observed during the turn-on. Therefore,
we used a partial covering model which combines both, scattering and
absorption, to fit the data over the time of the turn-on. As components for
the spectral model we used an exponentially cutoff power-law $I_{\rm
  power}(E)\cdot I_{\rm highecut}(E)$ as implemented in \texttt{XSPEC}, a
cyclotron line feature $I_{\rm cyc}(E)$ at 39\,\kev using the
\texttt{cyclabs} model of \texttt{XSPEC}, and a Gaussian emission line
$I_{\rm Fe}(E)$ fixed at 6.4\,\kev. For an analytic description of the
individual model components we refer the reader to the \texttt{XSPEC}
manual \citep[][]{arnaud:02a}.

\begin{figure}
  \resizebox{\hsize}{!}{\includegraphics{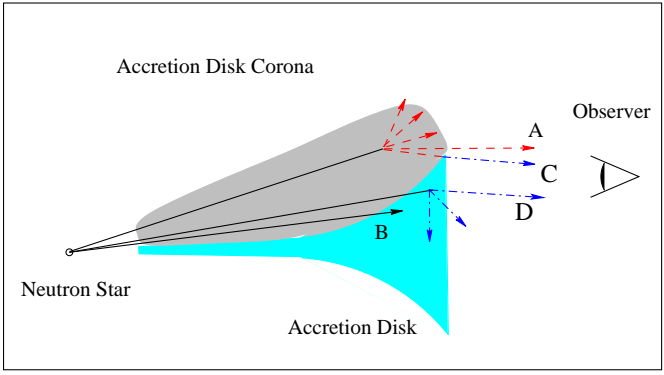}}
  \caption{Schematic illustration of the geometry during a
    turn-on of \her (not to scale). The accretion disk rim, the accretion
    disk corona, and the neutron star are shown. Three different components
    contributing to the overall observed flux are indicated: radiation
    absorbed by the cold material of the accretion disk (photons following
    beam~A), radiation absorbed by the accretion disk rim and further
    reduced in flux by Thomson scattering (photons following beam~B), and
    radiation scattered by the corona into the line of sight of the
    observer (photons following beam~C).
    \label{fig:disk-corona-model}}
\end{figure}

We combined these four spectral components to provide a basic continuum
model, which is then observed through an absorber partially covering the
continuum source.  In an analytic form the final spectral model describing
the photon spectrum can then be written as
\begin{align}
  I(E)&= B \cdot (1+a(E)) \cdot \underbrace{(I_{0}(E)+I_{\rm
  Fe}(E))}_{\mbox{Primary Spectrum}}
\intertext{with}
  a(E)&=\underbrace{e\,^{-N_{\rm  H}\sigma_{\rm
        T}} }_{\mbox{Thomson-Scattering}}\cdot\; C\;\cdot\underbrace
        { e\,^{-N_{\rm H}\sigma_{\rm bf}}
    }_{\mbox{Absorption}} \label{eqn:parta}
\intertext{and}
  {I_{0}(E)}&=I_{\rm power}(E)\;\cdot\;I_{\rm highecut}(E)\;\cdot\;I_{\rm
  cyc}(E)
\end{align}
where $\sigma_\text{T}$ is the Thomson cross-section and where the
bound-free absorption cross sections, $\sigma_\text{bf}$, are those used in
the \texttt{tbabs} model of \texttt{XSPEC} \citep[][]{wilms:00c}. For the
fitting the absorption and scattering part in Eq.~(\ref{eqn:parta}) uses
the same $N_\text{H}$. For the Thomson scattering part in
Eq.~(\ref{eqn:parta}) the electron density calculates from $N_\text{H}=
1.21\,N_\text{e}$, assuming material of solar abundances
\citep[][]{wilms:00c}. The constant $C$ in Eq.~(\ref{eqn:parta}) defines
the relative ratio of absorbed and scattered radiation to the unaffected
radiation.  A larger value of $C$ implies a larger degree of absorbed and
scattered flux. During the remainder of this paper the unaffected model
component will be called MC~I and the component influenced by absorption
and Thomson scattering will be referenced as MC~II. For the model component
MC~I we neglected Galactic absorption since the Galactic
$N_\text{H}=1.79\,10^{20}\,\text{cm}^{-2}$ is small compared to the
$N_\text{H}$ of model component MC~II. A schematic picture
  illustrating the geometric situation during the turn-on is shown in
  Fig.~\ref{fig:disk-corona-model}. The positions of the accretion disk and
  an accretion disk corona relative to the neutron star are shown. Photons
  following beam~A, marked by a dashed line, are scattered in the corona
  and can be partially absorbed in the accretion disk rim (beam~C, dash
  dotted line). While a certain fraction of photons are blocked by the
  accretion disk (beam~B), the outer parts of the accretion disk are
  optically thin for X-rays.  Therefore, the spectral distribution of
  photons following beam~D will show strong signature of photoelectric
  absorption. To simplify matters, photons reaching the observer directly
  without being modified by either the accretion disk nor the accretion
  disk corona are not shown in Fig.\ref{fig:disk-corona-model}. Comparing
  this geometric interpretation with the spectral model given in
  Eq.~\eqref{eqn:parta} allows the following interpretation:
\begin{itemize}
\item The unaffected model component $I(E)= I_{0}(E)+I_{\rm Fe}(E)$, called
  MC~I from now on, accounts for photons following beam~A and photons
  reaching the observer directly (this case is not shown in
  Fig.~\ref{fig:disk-corona-model}).
\item The model component modified by photoelectric absorption and Thomson
  scattering $I(E)= a(E) \cdot (I_{0}(E)+I_{\rm Fe}(E))$, called MC~II from
  now on, represents the spectral distribution of photons following beam~C
  and beam~D.
\end{itemize}
We emphasize that it is not possible to use a physically more realistic
spectral model which treats scattered, absorbed, and direct flux
separately. The problem lies within the nature of Thomson scattering: for
photon energies $E\lesssim 10\,\kev$ ($E \ll m_{\text{e}}c^2$) and when the
influence of multiple scatterings can be neglected ($\tau \lesssim 5$),
scattering of photons by stationary and free electrons can be treated as
elastic and consequently energy independent (classical Thomson
approximation). This makes it impossible to separate, e.g., the direct flux
and the flux contribution from photons scattered into the line of sight
(beam~A) using spectral analysis alone.
\begin{figure*}
  \centering
  \includegraphics[width=17cm]{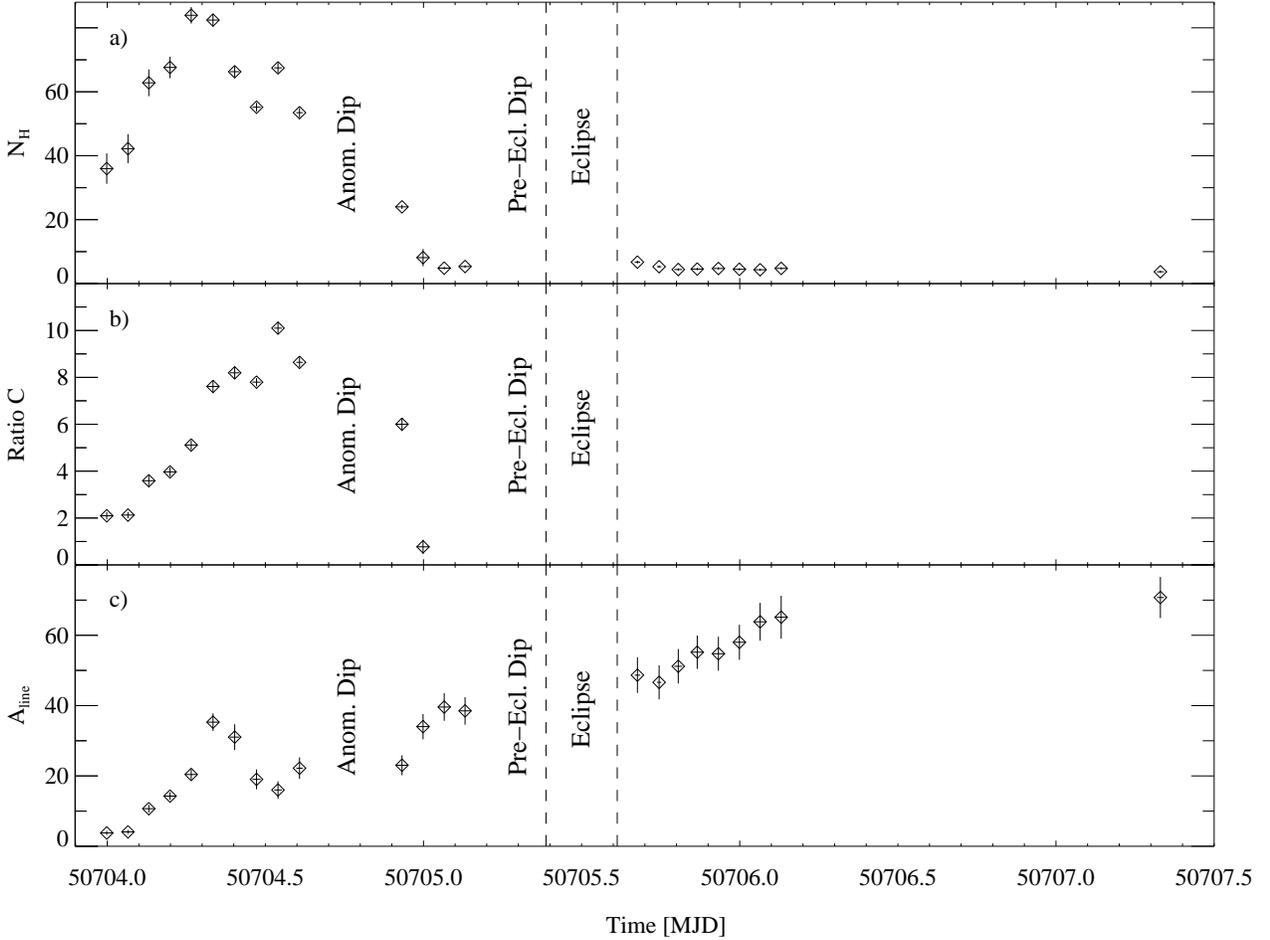}
  \caption{From top to bottom: (a) $N_\text{H}$ [$10^{22}\text{cm}^{-2}$],
    (b) ratio between the model component MC~II and MC~I as defined in
    Eq.~\eqref{eqn:parta}, and (c) normalization of the iron line
    $A_\text{line}$. The uncertainties are $\pm 1\sigma$.}
  \label{fig:parameter}
\end{figure*}

Using this simplified partial covering spectral model all 32 phase averaged
\pca and \hexte spectra were fitted in the energy range
$2.9$--$18\,\text{\kev}$ (\pca) and $15$--$100\,\text{\kev}$ (\hexte). In a
first iteration we fitted the data with all parameters free except the
power-law index $\alpha$ which was kept fixed at $1.068$. This is an
average value for $\alpha$ derived from the data with high counting
statistics towards the end of the turn-on (orbits~15--31). The results for
the remaining free fit parameters are listed in
Table~\ref{tab:results-spectral-analysis}. This analysis reveals that the
folding energy $E_\text{cutoff}$, $E_\text{fold}$, $E_\text{cyc}$,
$\sigma_\text{ cyc}$, $E_\text{Fe}$, and $\sigma_\text{Fe}$ show no
significant variation over the duration of the turn-on. Therefore, these
values were kept fixed at their mean values (see
Table~\ref{tab:fixedvalues}) for the further analysis. Leaving these values
fixed allows us to determine the variation of the remaining free parameters
for the time of turn-on, which are the neutral column density $N_\text{H}$,
the ratio $C$, the normalization of the power law $A_\text{PL}$, and of the
iron emission line $A_\text{Line}$.  The results are shown in
Fig.~\ref{fig:parameter} and the corresponding fit parameters are given in
Table~\ref{tab:her_short_fits}.

\begin{table}
  \caption{Parameters fixed to their mean value for the fitting of the
    spectra over the time of the turn-on.}\label{tab:fixedvalues}
  \centering
  \begin{tabular}{lr}\hline\hline
    Parameter             & Value      \\ \hline
    $\alpha$              & 1.068      \\ 
    $E_\text{cutoff}$     & 21.5\,\kev \\
    $E_\text{fold}$       & 14.1\,\kev \\
    $E_\text{cyc}$        & 39.4\,\kev \\
    $\sigma_\text{cyc}$   & 5.1\,\kev  \\ 
    $E_\text{Fe}$         & 6.45\,\kev \\
    $\sigma_\text{Fe}$    & 0.45\,\kev \\
    \hline
  \end{tabular}
\end{table}

\subsection{Spectral Parameters}\label{subsec:specevol}
Assuming a simple geometrical model of a turn-on where the outer edge of
the accretion disk opens the line of sight to the central neutron star as
indicated in Fig.~\ref{fig:disk-corona-model}, the neutral column density
$N_\text{H}$ is expected to gradually decrease when the flux increases. As
can be clearly seen from Fig.~\ref{fig:parameter}a, the behavior of
observed $N_\text{H}$ is contrary to this simple model: As we have already
mentioned earlier \citep{kuster:99a}, $N_\text{H}$ increases during the
first four \xte orbits, reaches a maximum during orbit~5, and then declines
until orbit~17 where it becomes untraceable. This evolution goes in
parallel to the degree of scattered and absorbed radiation, $C$, shown in
Fig.~\ref{fig:parameter}b, which increases until orbit~09 and afterwards
starts to decline. The turning point in the progression of $C$ matches
almost exactly the maximum of $N_\text{H}$. In contrast to $N_\text{H}$ and
$C$, the normalization of the Fe emission line, $A_\text{Line}$ reproduces
the progression of the count rate and increases during the whole of the
turn on (Fig.~\ref{fig:parameter}c). We will come back to the
interpretation of this behavior of $N_\text{H}$ and $C$ in
Sect.~\ref{sec:conclusion}.

% Only later on (later than orbit~05), $N_\text{H}$ mirrors the count
% rate indicating that photoelectric absorption is the cause for the
% changing amount of flux (cf.  Fig.~\ref{turnonlc} and
% Fig.~\ref{fig:parameter}c). Note that the turning point in the
% progression of $C$, which lies between orbit~4 and 5, matches almost
% exactly the maximum in $N_\text{H}$. In contrast to the observed
% $N_\text{H}$, the normalization of the iron emission line
% $A_\text{Line}$ reproduces the progression of the count rate (cf.\ 
% Fig.~\ref{turnonlc}).

\section{Evolution of the pulse profile}\label{sec:pulsevol}

\subsection{Pulse variation depending on disk phase}
Further insight into the physics of the turn-on comes from the substantial
changes in shape and amplitude of the X-ray pulse with 35 day phase. One of
the earliest studies of these changes was presented by \citet{bai:81a}, who
observed that the hard central peak and the soft trailing shoulder of the
\her pulse profile are affected differently over the time of the turn-off.
He interpreted this effect by a time dependent covering of the two polar
emission regions on the neutron stars surface by the inner accretion disk
rim and its corona.  Since the thickness of the inner accretion disk rim
and the neutron star are of the same order of magnitude, they have similar
angular size. Consequently, the observed flux from the two neutron star
poles can be affected differently in time by the material of the inner disk
rim.
\begin{figure}
  \resizebox{\hsize}{!}{\includegraphics{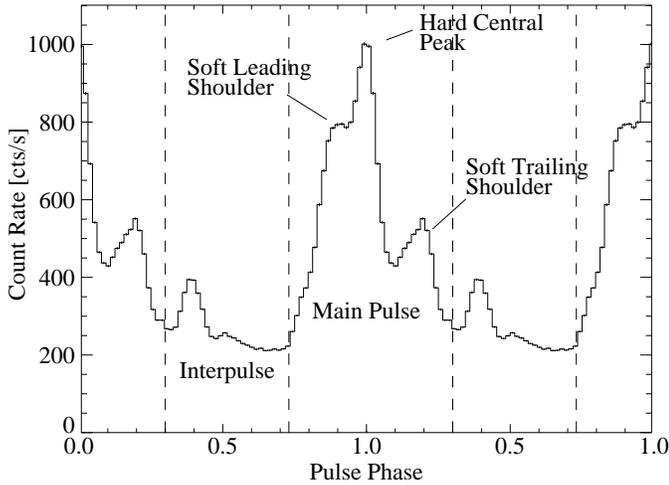}}
  \caption{Nomenclature for the different features of the \her pulse
    profile. This pulse profile is taken from orbit~30 in the energy range
    of 9.0--13.0\,\kev.}\label{fig:pulseprofile}
\end{figure}
Building upon this work and on observations of the change of the
pulse-profile during the end of the 35 day cycle, \citet{scott:00a} were
able to present a refined geometric model explaining these changes in more
detail.

\subsection{Observed pulse variation}
\begin{figure}
  \resizebox{\hsize}{!}{\includegraphics{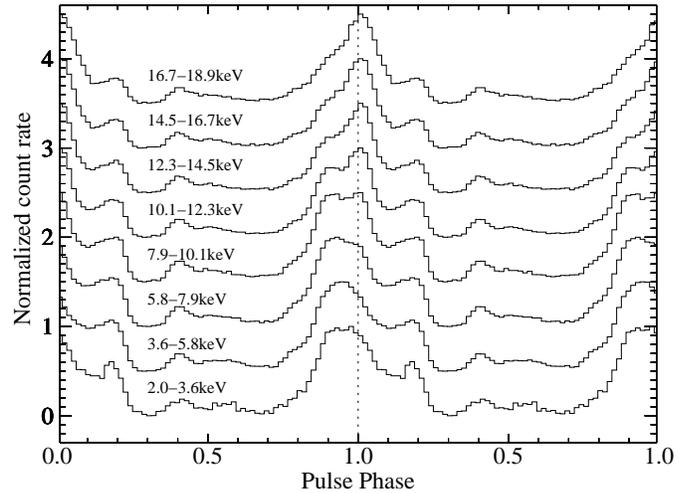}}
  \caption{Energy dependence of the pulse profile. The pulse profile of
    orbit~31 is shown in eight different energy bands. All profiles are
    normalized to unity. Each profile, except the profile of the energy
    range 2.0--3.6\,\kev, is shifted by 0.5 in the $y$ direction relative
    to the previous profile. Note, the energy dependant change of the
    relative intensity of the soft leading shoulder to the hard central
    peak, which results in a double peaked structure close to pulse phase
    1.0. This feature is most pronounced at energies between
    10--14\,\kev}.\label{fig:enerypuls}
\end{figure}

For the following discussion, we use the nomenclature given in
Fig.~\ref{fig:pulseprofile} for the various features of the \her pulse
profile. Fig.~\ref{fig:enerypuls} shows how the relative strength of these
features change with energy. The most distinct variation apparent, is the
decreasing flux of the soft leading shoulder with increasing energy. This
change leads to a double peaked structure consisting of the hard central
peak and the soft leading shoulder that clearly can be identified in the
energy range 6--12\,\kev \citep[][]{deeter:98a}.

\begin{figure*}
  \centering
  \includegraphics[width=17cm]{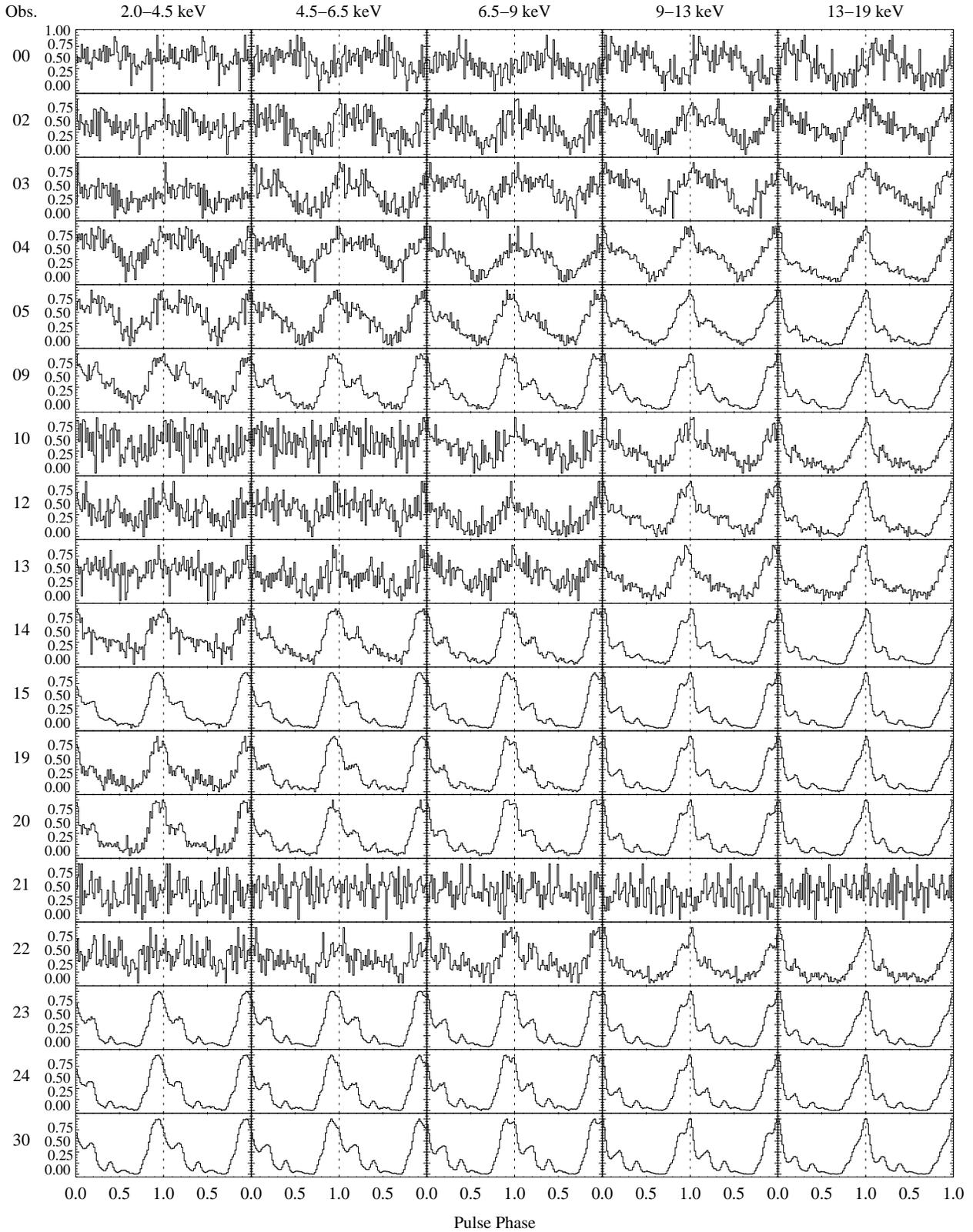}
  \caption{Evolution of the pulse profile as observed over the time of the
    turn-on. All profiles are normalized to unity at the maximum of the
    main pulse, after subtraction of the off-pulse constant flux. Pulse
    phase 0 is defined as the maximum of the main pulse in the energy band
    of 13--19\,\kev.}
  \label{fig:pulspanel}
\end{figure*}

During the early phases of the turn-on, strong photoelectric absorption and
Thomson scattering will modify the pulse shape (Fig.~\ref{fig:pulspanel}).
Thomson scattering in a hot plasma causes the broadening of the pulse
profile in all energy bands which leads to an almost sinusoidal pulse shape
during early times of the turn-on (orbits~00--05).  During later phases of
the turn-on (orbits~05--09) only the lowest energy channels are affected by
strong noise. This implies that during this phase energy dependant
photoelectric absorption is the dominant process.  The behavior during
orbits~05--09 is very similar to the situation in orbits~10--14 during
which the anomalous dip took place, which is presumably caused by cold
material located at the outer rim of the accretion disk, crossing the line
of sight to the neutron star \citep[][]{shakura:99a}. During the eclipse
phase (orbit~21--22) all energy channels are affected by strong noise and
no pulsation was detected.
\begin{figure*}
  \centering
  \includegraphics[width=17cm]{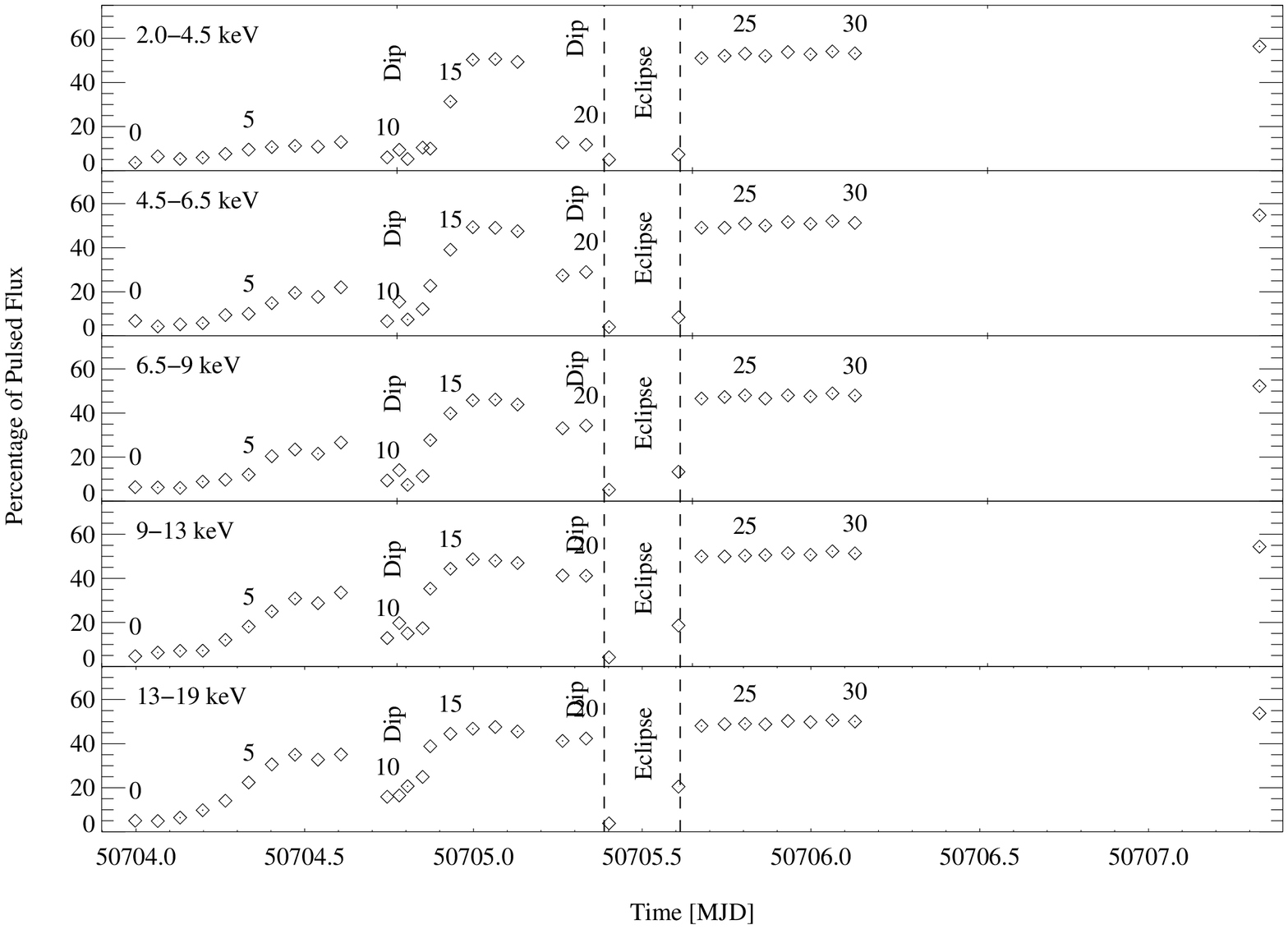}
  \caption{Pulsed fraction over the time of the turn-on in different energy
    ranges. To determine the pulsed fraction the pulse profiles have to be
    smoothed to reduce Poisson noise. The pulsed fraction is given as
    percentage of the pulsed flux compared to the non pulsed flux.
    \label{fig:pulsedfraction}}
\end{figure*}

These effects of scattering and photoelectric absorption can also be seen
in the behavior of the pulsed fraction with time
(Fig.~\ref{fig:pulsedfraction}). It is clearly visible that $F_\text{
  pulsed}$ increases more rapidly in the high energy bands. At lower
energies, the pulsed flux is suppressed towards the beginning of the
turn-on, similar to the situation observed during egress of the anomalous
dip, where the pulsed flux increases faster at higher energies.  At the end
of the turn-on, after orbit~23, the pulsed fraction is almost constant.
Contrary to later phases of the main-on and the turn-off the intrinsic
pulse shape does not change significantly over the time of the turn-on.
This result is in agreement with earlier findings of, e.g.,
\cite{gruber:80a}, \cite{bai:81a}, \cite{truemper:86a}, or
\cite{deeter:98a}. The pulse profile observed at the beginning of the
turn-on can be interpreted, therefore, as a main-on pulse profile modified
by the influence of photoelectric absorption and scattering.

\section{Simulating pulse variation}\label{sec:pulsesim}
To quantify the effects of scattering and photoelectric absorption on the
change of the X-ray spectrum and the pulse profile shown in
Sects.~\ref{sec:specevol} and~\ref{sec:pulsevol}, we now turn to Monte
Carlo simulations of the radiation transport of the pulsar's flux in a
scattering medium. For the simulations we assume that the pulse shape seen
at the end of the turn-on (orbit~30), is the intrinsic pulse shape caused
by the emission characteristic of the neutron star and is not changing over
the time of the turn-on. Furthermore, we assume that the smearing of the
pulse profile and the change in spectral shape at the beginning of the
turn-on are solely caused by scattering and photoelectric absorption in the
medium covering the line of sight to the neutron star. With these
assumptions we can use the pulse profile observed in orbit~30 as a
``template'' profile and investigate the effects of a scattering and
absorbing corona on the pulse shape depending on $N_\text{H}$ and the size
of the scattering region. This is done via Monte Carlo simulations, as
described in the following sections.  
 
\subsection{Monte Carlo Simulations}
Assuming a point source emitting the intensity $I(t_0)$ at time $t_0$, the
intensity at infinity observed at time $t$ can be written as
\begin{equation}\label{eqn:green}
I^{\infty}(t)=\int_{-\infty}^{t} G(t,t_0)I(t_0)dt_0
\end{equation}
where $G(t,t_0)$ is the scattering Green's function, i.e., the
appropriately normalized solution of the time-dependent equation of
radiation transfer through the scattering and absorbing
medium for a $\delta$-function pulse of light emitted at time
$t_0$. 

First analytical approaches to determine the Green's function for
scattering in a cold corona were based on the fundamental method
developed by \citet{lightman:81a}. \citet{brainerd:87a},
\citet{kylafis:87a}, and \citet{kylafis:89a} found analytical
solutions for $G(t,t_0)$ for simple geometries, such as the diffusion
of photons out of a spherical shell surrounding a central point
source. For the general case, analytical results are difficult to
obtain and one has to resort to numerical solutions instead. Here, we
use a modification of the linear Monte Carlo code based on the method
of weights \citep{sobol:91a} that we have previously used to compute
$G(t,t_0)$ for the case of a hot Comptonizing plasma
\citep{nowak:99a}. In our simulations we consider Compton scattering
(using the Klein-Nishina cross section), photoelectric absorption from
material of solar abundance using the cross sections of
\citet{verner:96a}, and fluorescent line emission \citep[using the
fluorescence yields of][]{kaastra:93a}. We model the propagation of
photons through a plane-parallel slab with thickness $d$ and hydrogen
column $N_\text{H}$ (corresponding to a certain optical depth for
electron scattering, $\tau_\text{es}$) that is illuminated by a source
at infinity.  We consider both, neutral and fully ionized slabs.
Output of the simulation is $G(t,t_0)$ as a function of $N_\text{H}$,
$d$, and energy band and the angle-dependent photon spectrum of
photons leaving the slab. We normalize the time to the light crossing
time of the slab.
\begin{figure}
  \resizebox{\hsize}{!}{\includegraphics{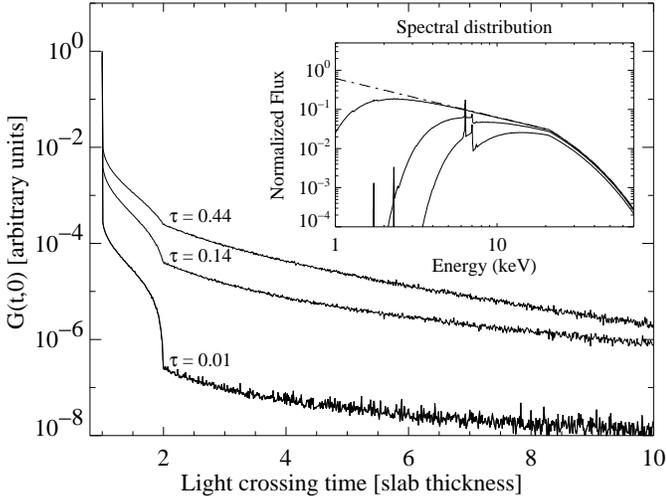}}
  \caption{$G(t,t_0)$ in the energy range 1.0--70.0\,\kev for a neutral
    corona and low optical depths of $\tau_\text{es}=0.01$
    ($N_\text{H}=1.0\,10^{22}\;\text{cm}^{-2}$), $\tau_\text{es}=0.14$
    ($N_\text{ H}=1.25\,10^{23}\;\text{cm}^{-2}$), and
    $\tau_\text{es}=0.44$ ($N_\text{H}=2.75\,10^{24}\;\text{
      cm}^{-2}$). The peak at $t=1$ is caused by photons crossing the
    slab without scattering, the break at $t=2$ is caused by photons
    scattering at most one time before leaving the slab.
    \emph{Inset}: Spectrum emerging from the slab for the same
    columns. Note the emergent fluorescent emission lines for the
    larger optical depths.
    \label{fig:greensabs}}
\end{figure}

\begin{figure}
  \resizebox{\hsize}{!}{\includegraphics{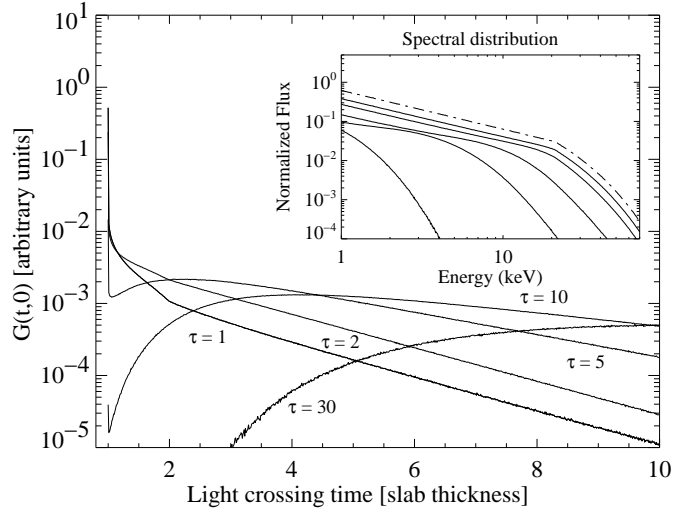}}
  \caption{$G(t,t_0)$ for the same spectral distribution as shown in
    Fig.~\ref{fig:greensabs} but for a fully ionized medium and higher
    optical depths.
    \label{fig:greension}}
\end{figure}

As an example, Fig.~\ref{fig:greensabs} shows $G(t,t_0)$ for a neutral slab
with $N_\text{H}=1.0\,10^{22}\,\text{ cm}^{-2}$,
$1.25\,10^{23}\;\text{cm}^{-2}$, and $2.75\,10^{24}\;\text{ cm}^{-2}$,
while Fig.~\ref{fig:greension} shows the same for a fully ionized slab.
For $\tau_\text{es} \lesssim 5$ the Green's function $G(t,t_0)$ in
Fig.~\ref{fig:greensabs} and Fig.~\ref{fig:greension} is dominated by
photons crossing the slab on a straight line without being scattered from
electrons. These photons contribute to the peak apparent at diffusion times
equal to the light crossing time of the slab ($t_\text{diff}=1$) and only a
small number of photons diffuse out of the slab after this initial peak.
For increasing electron optical depth, the number of diffusing photons
increases significantly since the mean number of scatterings per photon
increases approximately as $\sim$$\tau_\text{es}^2$.  As a consequence, the
maximum of $G(t,t_0)$ moves towards higher diffusion times, until
$G(t,t_0)$ is dominated by photons scattered multiple times
($\tau_\text{es}>5$). The energy loss per electron scattering event can be
estimated for a $15\,\text{\kev}$ photon to $\Delta E \approx
0.44\,\text{\kev}$. This implies that, as soon as the mean number of
scattering per photon increases above $\sim$15--20 ($\tau_\text{es} \gtrsim
4$), photons of the energy range between 10--20\,\kev are redistributed to
the energy band of 2--10\,\kev. This redistribution effect is the origin of
the changing spectral shape with increasing optical depth visible in
Fig.~\ref{fig:greension} (inset).  As a consequence, the cut-off energy
moves towards lower energies and a bump at energies between 1--5\,\kev
arises. For even larger optical depths, $\tau_\text{es}>10$, this bump
slowly vanishes and the flux above 6\,\kev decreases rapidly. The overall
spectral shape is then given by the left most spectrum shown in the top
panel of Fig.~\ref{fig:greension}.

Considering a neutral medium, absorption plays the dominant role over the
temporal effects of Compton scattering. For a neutral medium with
$\tau_\text{es}\gtrsim 1$, almost all flux below 10\,\kev is
suppressed\footnote{For material with solar abundance, the photoionization
  cross section equals the Thomson cross section at $\sim$10\,\kev. Due to
  its $E^{-3}$ proportionality the optical depth below this energy will
  therefore be always higher than the electron optical depth.}.  At such
optical depths the effects caused by diffusion time are still negligible
since the mean number of scatterings per photon is close to unity. Thus, to
achieve noticeable changes in beam shape a high fraction of ionized
material is needed.  On the other hand a high fraction of neutral material
only weakly alters the beam shape but reduces the flux at low energies.
\begin{figure}
  \resizebox{\hsize}{!}{\includegraphics{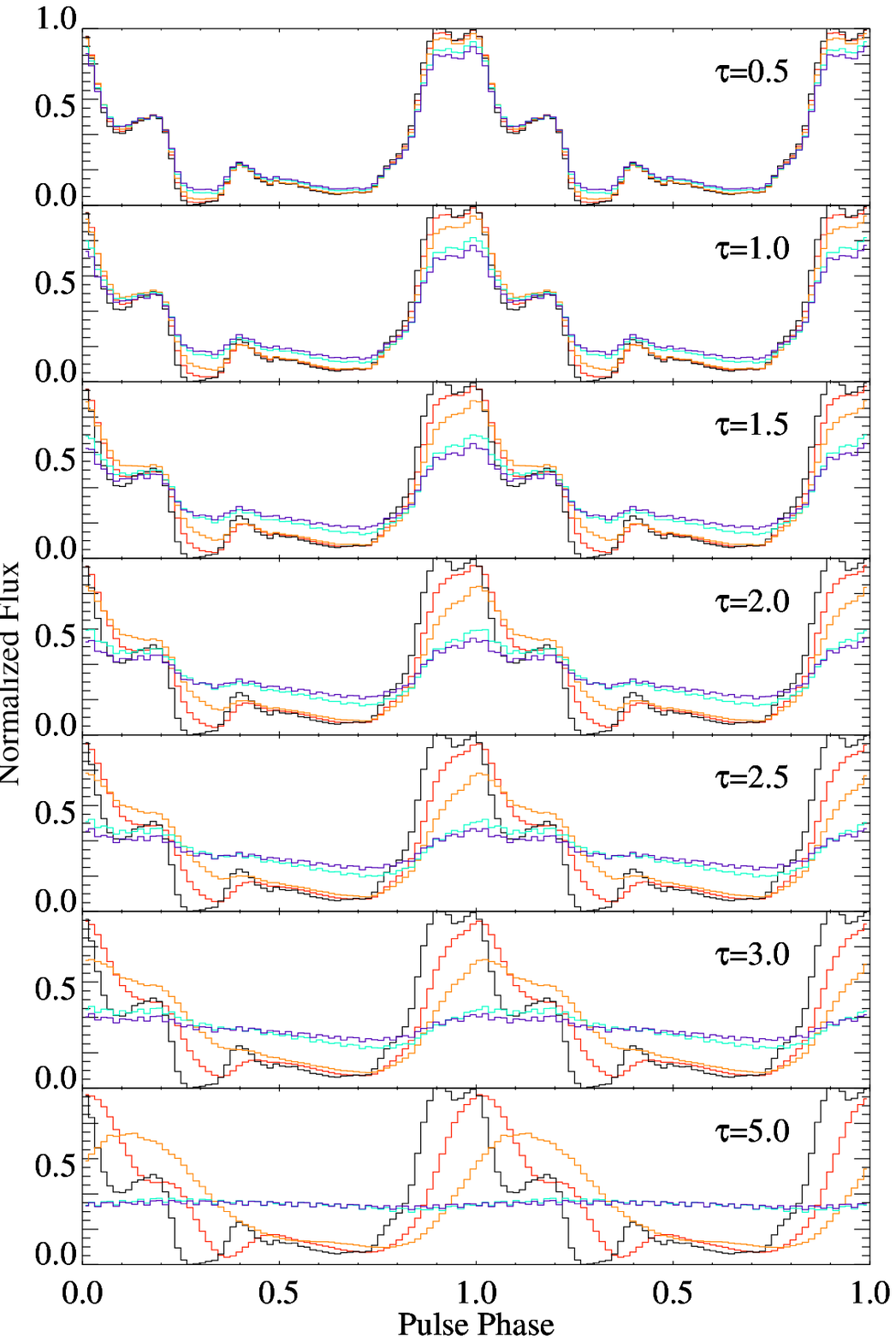}}
  \caption{Her X-1 pulse profile of orbit~30 (black) transmitted by a fully
    ionized scattering medium. The electron optical depth varies from
    top to bottom from $\tau_\text{es}=0.5$ to $\tau_\text{es}=5.0$.
    In addition, for each optical depth value the thickness $d$ of the
    scattering medium is set to $0.02$ (red), $0.02$ (yellow), $0.2$
    (blue), and $0.4$ times (dark blue) the light crossing time of the
    slab.\label{fig:pulsesimul}}
\end{figure}

Fig.~\ref{fig:pulsesimul} demonstrates the influence of a fully
ionized scattering medium on the pulse shape of Her X-1 for different
values of $\tau_{es}$ and a variable thickness of the slab.  It is
obvious that for large optical depth values ($\tau_\text{es} > 3$),
even a thin scattering layer ($d < 0.1$) is sufficient to completely
hide the pulse. In addition the structure of the pulse profile is
steadily ``washed-out'' to an almost sinusoidal pulse shape. For
$\tau_\text{es} < 1.5$ the profile's substructure (soft trailing
shoulder) is still apparent even for high values of $d$. Note
especially the shift in pulse phase, which is pronounced in the
modified pulse profile for $\tau_\text{es} > 2.0$.  This phase shift
corresponds to the shift of the maximum of $G(t,t_{0})$ apparent in
Fig.~\ref{fig:greension}.
\begin{figure*}
  \begin{minipage}{0.49\textwidth}
    \centerline{Orbit~05}
    \includegraphics[width=8.8cm]{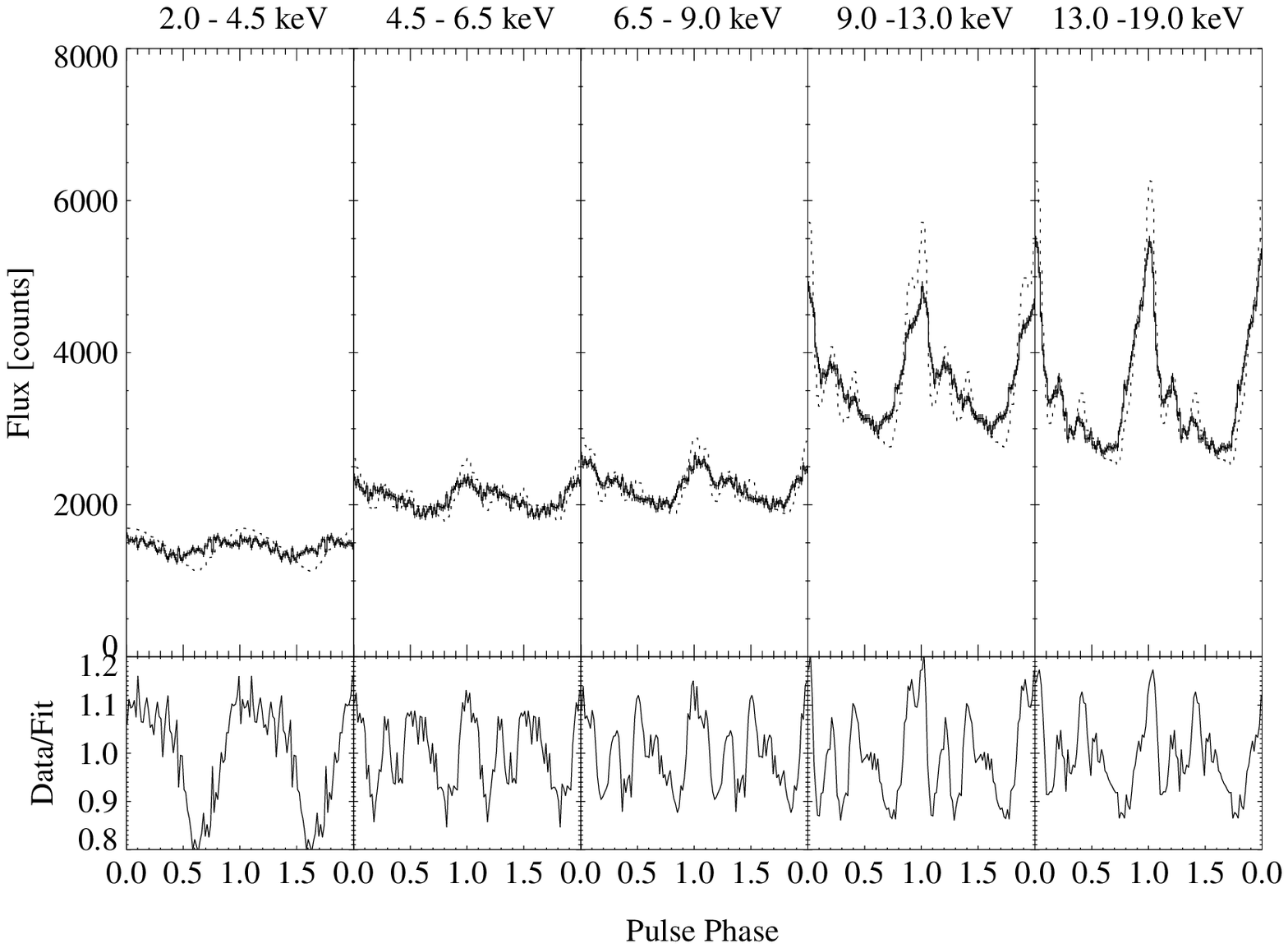}
    \vspace{0.2cm}
  \end{minipage}
  \begin{minipage}{0.49\textwidth}
    \centerline{Orbit~06}
    \includegraphics[width=8.8cm]{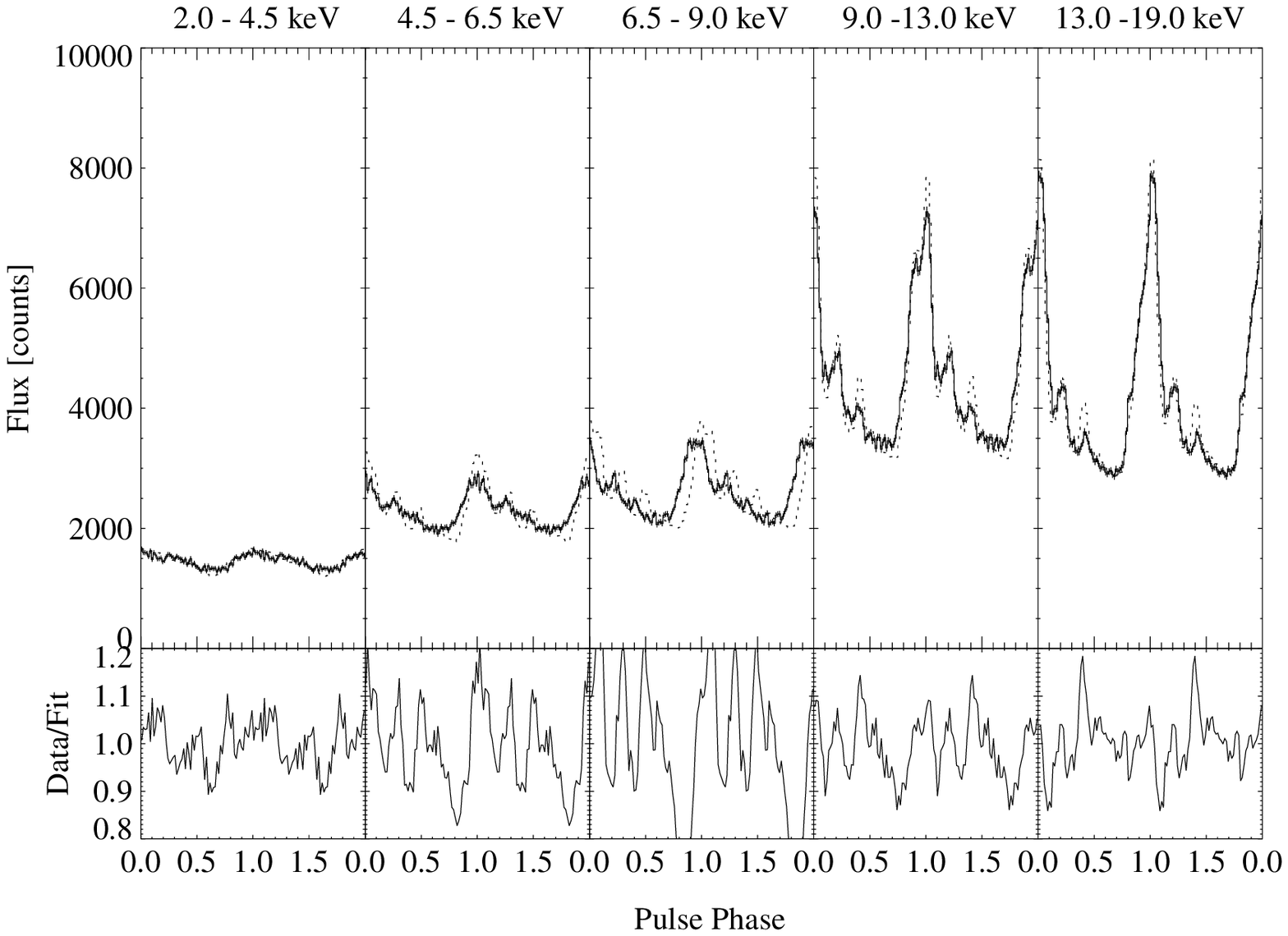}
    \vspace{0.2cm}
  \end{minipage}
  \begin{minipage}{0.49\textwidth}
    \centerline{Orbit~08}
    \includegraphics[width=8.8cm]{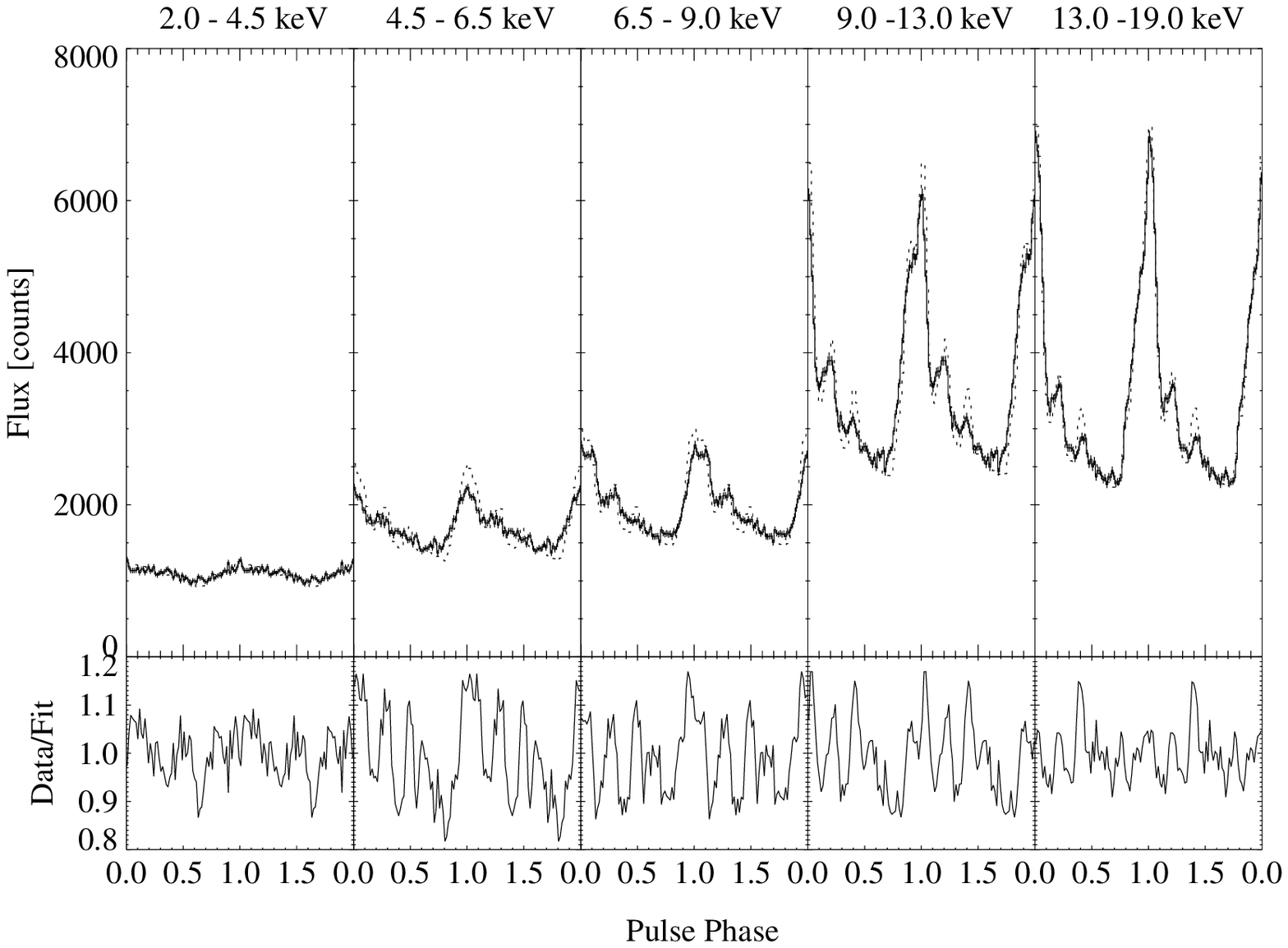}
    \vspace{0.2cm}
  \end{minipage}
  \begin{minipage}{0.49\textwidth}
    \centerline{Orbit~09}
    \includegraphics[width=8.8cm]{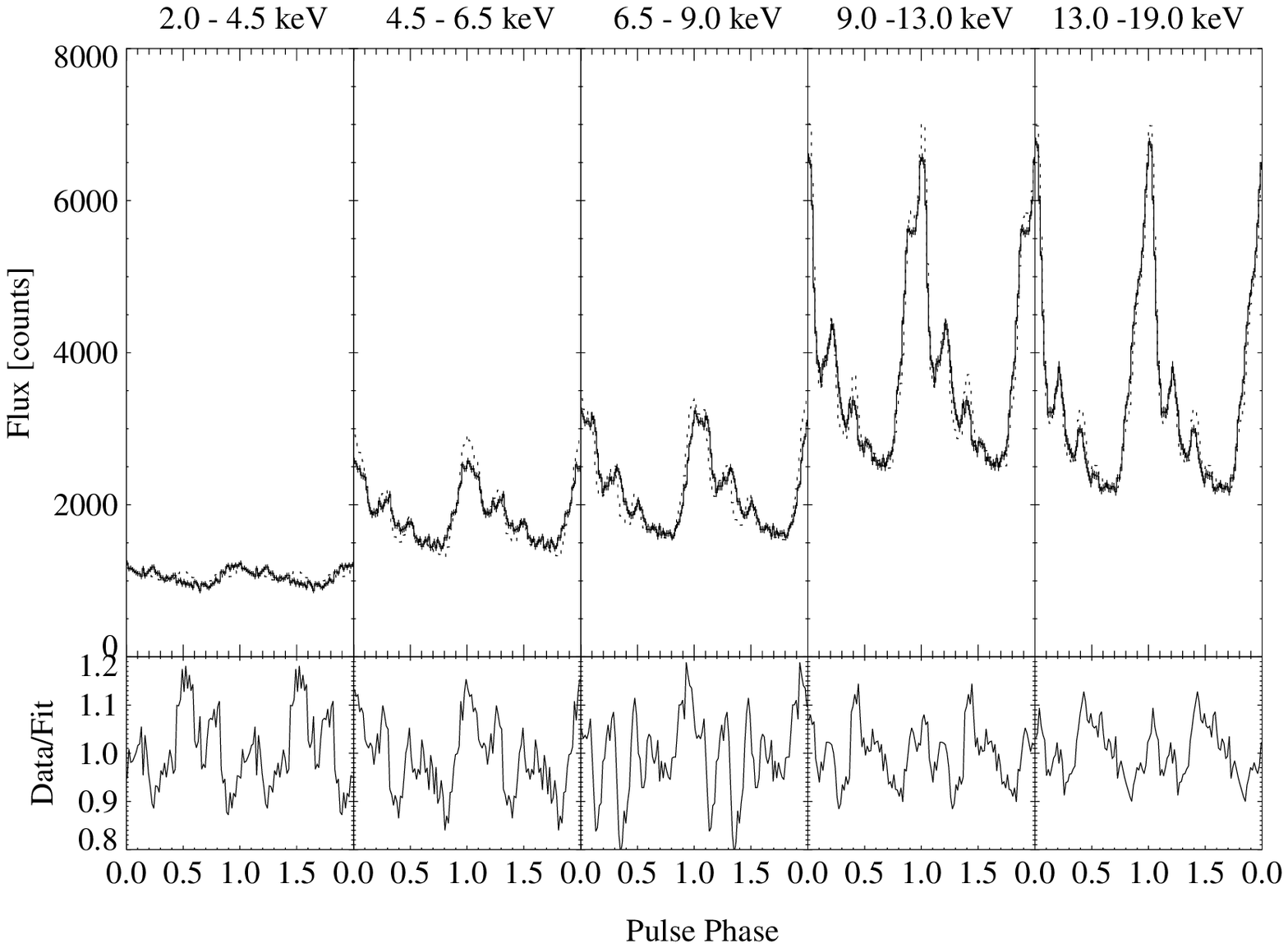}
    \vspace{0.2cm}
  \end{minipage}
  \begin{minipage}{0.49\textwidth}
    \centerline{Orbit~15}
    \includegraphics[width=8.8cm]{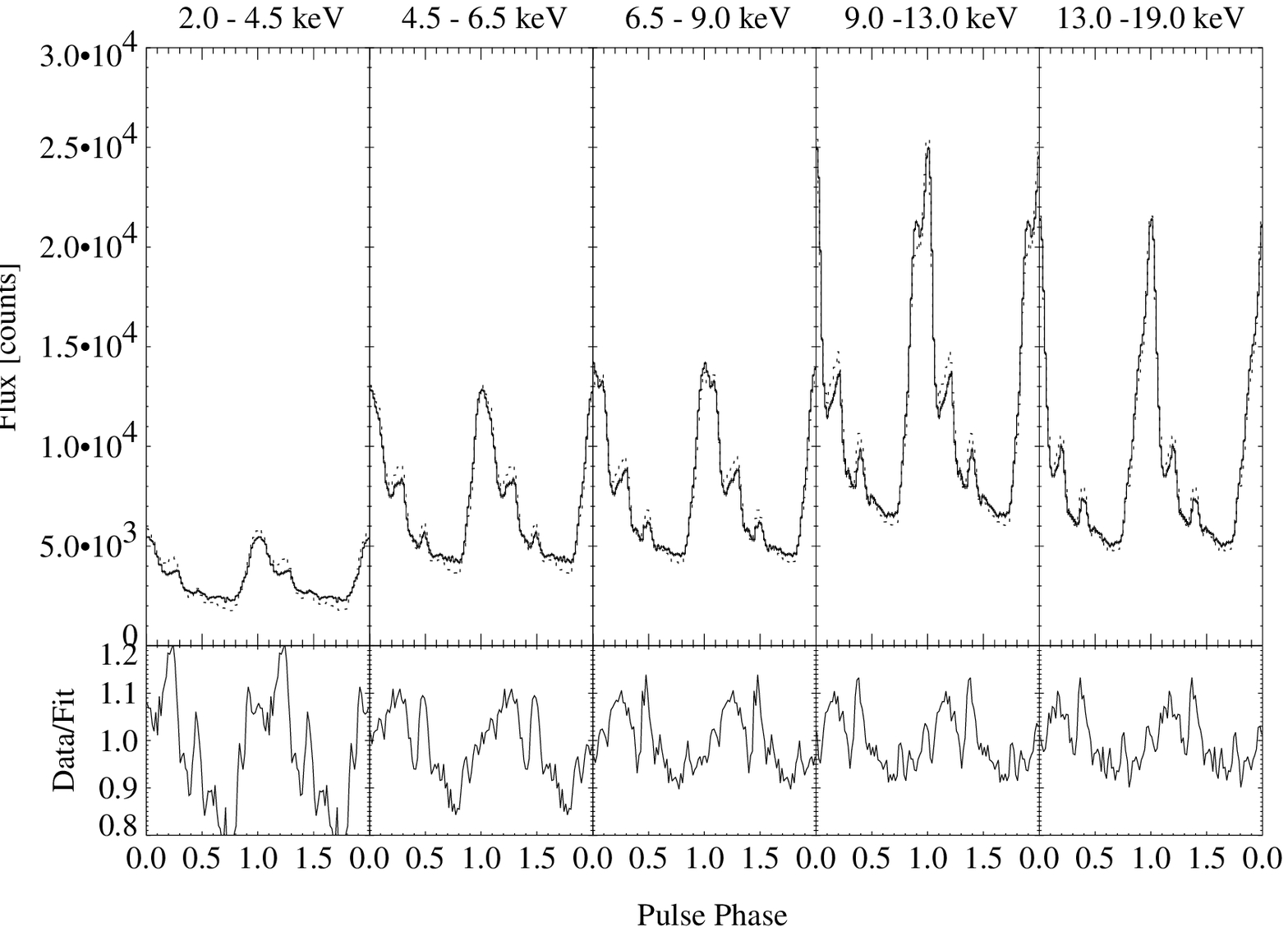}
  \end{minipage}
  \begin{minipage}{0.49\textwidth}
    \centerline{Orbit~24}
    \includegraphics[width=8.8cm]{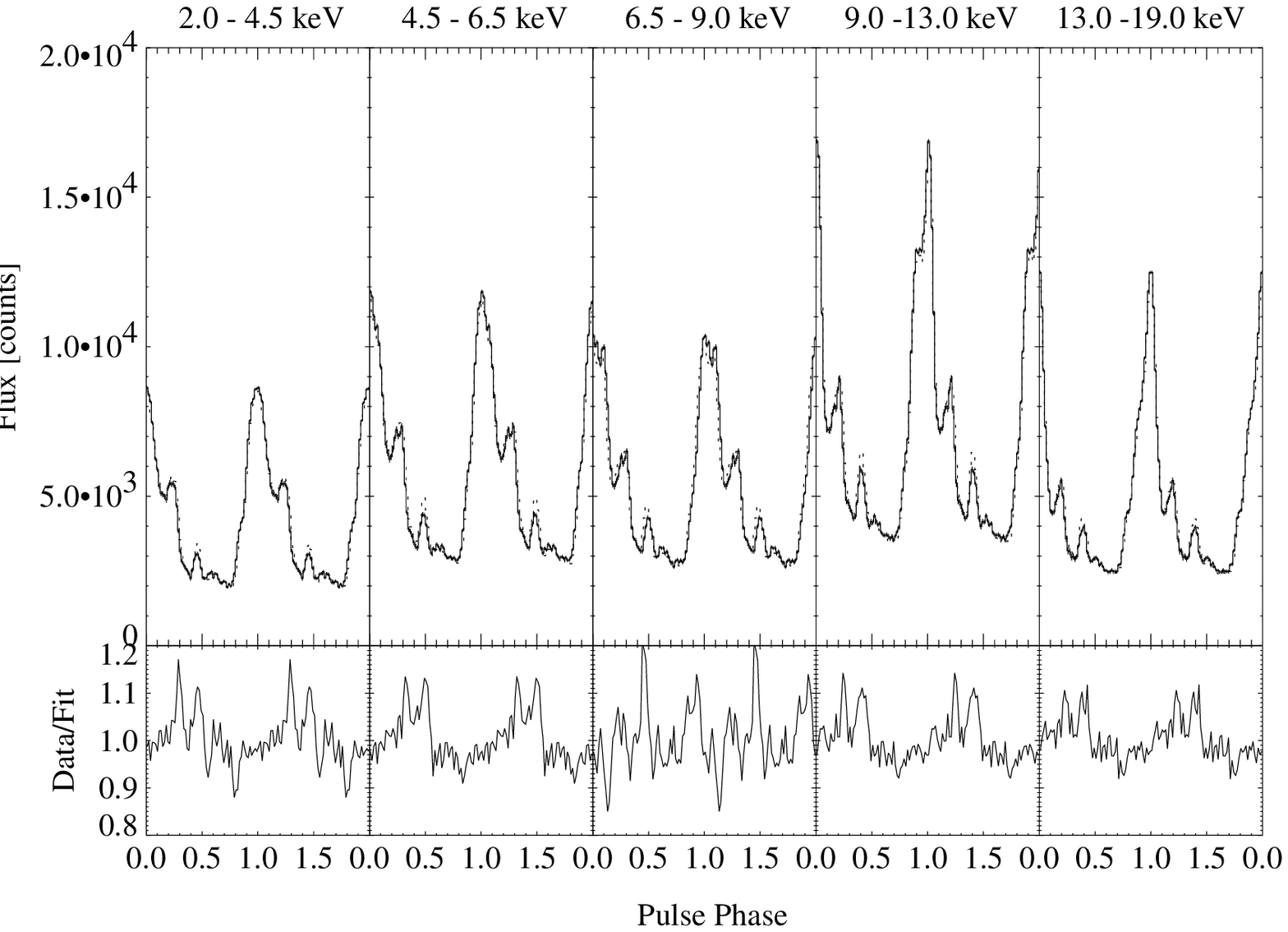}
  \end{minipage}
  \caption{Fit of \emph{simulated} pulse profiles to \emph{observed} pulse
    profiles for the selected observations (compare
    Fig.~\ref{fig:pulspanel}).  The solid line represents the count rate of
    the observed pulse profile of the indicated orbit, and the dashed line
    represents the simulated best-fit pulse profile. The residuals of the
    fits are given at the bottom. All energy bands are fitted
    simultaneously.
    \label{fig:simul-pulsefit}}
\end{figure*}

\subsection{Analysis of the Pulse Profile}
To simulate the variation of the pulse profile during the turn-on we
calculated $G(t,t_0)$ for the same energy ranges we used in
Section~\ref{sec:pulsevol} and each \xte orbit for two cases:
\begin{enumerate}
\item $G_\text{n}(t,t_0)$ for a neutral corona, with $N_\text{H,n}
  \approx N_\text{H}$ of the spectral analysis for each single orbit.
\item $G_\text{ion}(t,t_0)$ for a completely ionized corona with
  $1.0\,10^{22}\,\text{cm}^{-2} \le N_\text{H,es} \le
  9.5\,10^{24}\,\text{cm}^{-2}$.
\end{enumerate}
The Green's functions $G_\text{n}(t,t_0)$ and $G_\text{ion}(t,t_0)$ can
then be compared directly with the spectral model given by
Eq.~\eqref{eqn:parta}. Following the notation introduced in
Sect.~\ref{subsec:model}, $G_\text{ion}(t,t_0)$ corresponds to the model
component MC~I and $G_\text{n}(t,t_0)$ to the model component MC~II used
for the spectral analysis. For a medium where a fraction $f$ is fully
ionized, the total Green's function is given by $G(t,t_0)=(1-f)\,
G_\text{ion}(t,t_0)+ f G_\text{n}(t,t_0)$. Using a fixed
$G_\text{n}(t,t_0)$ with the respective $N_\text{H,n}$ for each orbit from
the spectral analysis and variable $G_\text{ion}(t,t_0)$, we can simulate
pulse profiles an observer located at infinity would see, by applying
Eq.~\eqref{eqn:green} to the ``template'' pulse profile.  As mentioned
above, the time scale of the simulated light curves is normalized to the
thickness of the corona, i.e., $t-t_0$ is measured in units of the light
crossing time of the slab.  Therefore, Green's functions for different
values of $d$ can be obtained by simply rescaling the time. For our
analysis we chose $d$ between 0.1 and 6 light seconds, appropriate for the
dimension of the accretion disk in \her with $r_\text{in} \sim 10^{8}$\,cm
and $r_\text{out} \sim 10^{11}$\,cm \citep[][]{cheng:95a}. To obtain a
proper flux normalization, the integrated flux of the profile of orbit~30
was set to unity. All other simulated pulse profiles are normalized
relative to this flux. From the spectral fitting of the partial covering
model to the observed spectra, the parameters of the spectral model
components MC~I and MC~II are known. By integrating the differential photon
flux of the model spectra over the specific energy bands of
Fig.~\ref{fig:pulspanel}, the total flux per energy range can be
calculated. Using the integrated photon flux both Green's functions,
$G_\text{n}(t,t_0)$ and $G_\text{ion}(t,t_0)$, can then be normalized to
the observed flux according to
\begin{equation}
  \label{eq:shot-normalization}
  \int_{t_\text{0}}^{t_\text{1}}\int_{0}^{\infty}R(h,E)N_\text{ph}(E,t)\,dE\,dt=\int_{-\infty}^{t}
  G_{E_{1},E_{2}}(t,t')\,dt 
\end{equation}
where $N_{\text{ph}}(E)$ is the differential photon flux and $R(h,E)$ the
detector response matrix.  Finally we performed a $\chi^2$ minimization fit
of \emph{simulated} to \emph{observed} pulse profiles as shown in
Fig.~\ref{fig:pulspanel} for each single \xte orbit.  The different energy
ranges were fitted simultaneously. This procedure allows us to determine
$N_\text{H,es}$, and $d$ as well as their uncertainty.
\begin{figure}
  \centerline{
    \includegraphics[width=88mm]{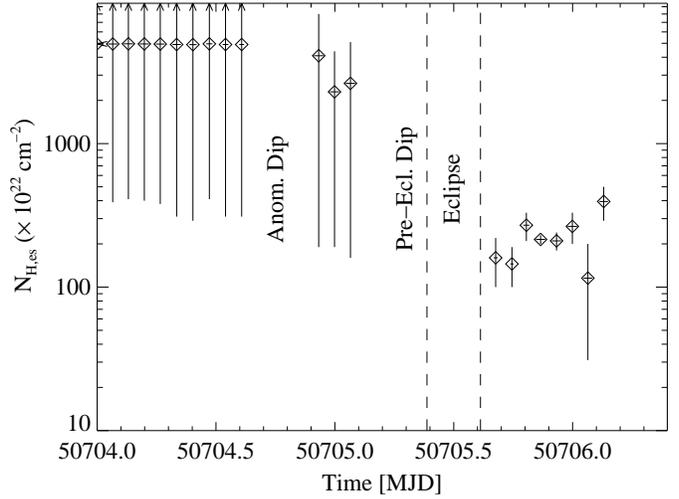}}
  \caption{$N_\text{H,es}$ as a function of time 
    determined with the method of pulse profile fitting as described in the
    text. The error bars are at the $1\sigma$ confidence
    level.}\label{fig:pulsenh}
\end{figure}

As an example, the results of the fit of \emph{simulated} to
\emph{observed} pulses are shown in Fig.~\ref{fig:simul-pulsefit} for
different orbits. The result for the selected orbit demonstrates, that
qualitatively and quantitatively the simulation can well reproduce the
observed pulses in all energy bands.  Similar results can be achieved for
observations not shown in Fig.~\ref{fig:simul-pulsefit}. The integrated
absolute flux in all energy bands is reproduced with an accuracy $\lesssim
10\%$. For observations earlier than orbit~04 the analysis suffers from the
low signal to noise ratio in the observed spectra and a resulting high
uncertainty in the normalization of $G_\text{ion}(t,t_{0})$ and $G_\text{
  n}(t,t_{0})$.

Fig.~\ref{fig:pulsenh} shows the overall development of the electron
optical depth over the full turn on. In the figure we give $\tau_\text{es}$
in terms of the electron column depth, $N_\text{H,es}$, to enable a direct
comparison with the results of the spectral fitting. During the early
phases of the turn on, $N_\text{H,es}$ is very high and the absorbing
medium is Compton thick. For the later observations, $N_\text{H,es}$
decreases and then levels out at a constant level. These results are a
qualitative confirmation of the behavior already expected from the visual
inspection of the pulse profile evolution: during the early stages the
pulse profile is completely dominated by scattering, while in later stages
the direct radiation becomes more and more important.  We note, however,
that the overall electron column deduced from our Monte Carlo simulations
is significantly higher than the column deduced from spectral fitting.
Given the explorative character of the Monte Carlo simulations and the
simplifications introduced in the model, this is not unexpected. For
example, our assumption of the scattering medium being a mixture of either
completely neutral or completely ionized is certainly an
oversimplification, as is the assumption of a slab geometry. On the other
hand, as we will show below, despite these simplifications the overall
trend of $N_\text{H,es}$ is in agreement with the common models for the
35\,day turn-on and thus reduces the ambiguities from the spectral
decomposition.

\section{Conclusion and discussion}\label{sec:conclusion}
In this paper we have shown that the evolution of the X-ray spectrum and
pulse profile during the 35\,day turn-on of Her~X-1 can be explained by
invoking a varying contribution of scattered and heavily absorbed photons
to the observed data (Sect.~\ref{sec:specevol} and~\ref{sec:pulsesim}).
Using Monte Carlo simulations, in Sect.~\ref{sec:pulsesim} we showed that
the observed behavior of these components appears to be consistent with the
results of pulse profile analysis with theoretical Green's functions for
the scattering and photoelectric absorption in the accretion disk. Despite
the existing limitations, such as modeling the accretion disk rim by a
simple slab with uniform density or modeling the scatterer as either fully
ionized or neutral, the methods applied here show that the distinct
contributions to the final pulse profile and spectrum can in principle be
separated by making use of the smearing of the pulse profile caused by the
scattering in the plasma. Our results confirm earlier work based on X-ray
spectral analysis alone on the nature of the turn-on of the 35\,day cycle
and on the nature of the accretion disk of Her~X-1
\citep[][]{davison:77a,parmar:80a,becker:77a,burwitz:01}. Our analysis
yields an optical depth of $\tau_\text{es}\approx3$--$10$ for the
scattering medium which is necessary to explain the observed smearing of
the pulse profile.  Such optical depths are consistent with coronal models
of neutron stars in LMXRBs \citep[e.g.,][and references
therein]{miller:00a}.

\begin{figure*}
  \centering
  \includegraphics[width=17cm]{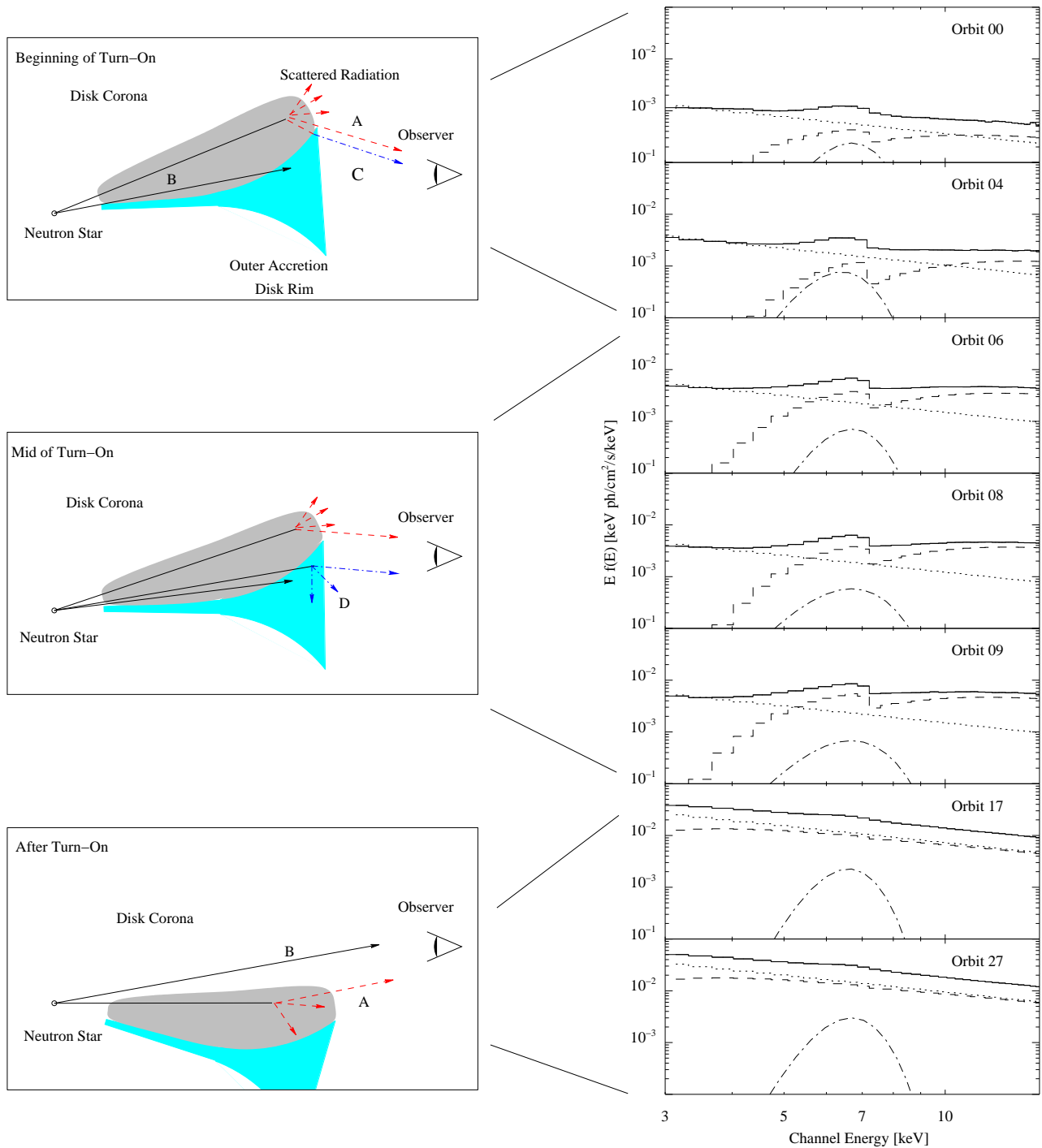}
  \caption{Schematic view of the outer accretion disk rim (blue),
    the accretion disk corona (grey), and the neutron star for three
    different times during the turn-on (not to scale). Right panel:
    Unfolded \pca spectra and the components of the spectral model for
    selected orbits.  The following components are shown: observed spectrum
    (solid line), spectral model component MC~I (dotted line) and MC~II
    (dashed line), and the Fe line.}
  \label{fig:turnonmodel}
\end{figure*}

The behavior summarized above can be explained by a simple geometric
model that takes into account the outer rim of an accretion disk that
opens the line of sight to the neutron star and the influence of a hot
accretion disk corona sandwiching the accretion disk.
Fig.~\ref{fig:turnonmodel} shows the positions of the accretion disk,
the disk corona, and the location of the observer are shown for
different times of the turn-on. The observed \pca spectra
corresponding to the phases of the turn-on and the unfolded spectral
model components for the energy range $3 \le E \le 12\,\text{\kev}$
are shown in the right panel of Fig.~\ref{fig:turnonmodel}. The
individual components of the spectral model are shown separately.

At the start of the turn-on, Thomson scattered radiation from the corona,
which is partially absorbed, dominates the observed spectrum.  The topmost
image of Fig.~\ref{fig:turnonmodel} ``Beginning of turn-on'' illustrates
the orientation of the disk and the location of the observer relative to
the disk for this time. Photons following beam~A, marked by a dashed line,
are scattered in the corona and partially absorbed in the disk rim
(beam~C). The direct view to the neutron star at this time is still blocked
and photons following beam~B cannot reach the observer.  The corresponding
spectra are those observed in orbits~00 and~04.  Since the scattered
spectral model component MC~I dominates the observed spectral flux, only a
small fraction of absorbed flux can be detected.  \emph{As a result, the
  total observed spectrum, which is the sum of all model components, shows
  only a weak signature of absorption and low values of $N_\text{H}$ are
  measured.} On the other hand, however, those scattered photons which
finally reach the observer have undergone many scattering events such that
the pulse profile is heavily distorted until orbit~06. In the pulse profile
fitting, this phase of the turn-on will therefore be characterized by large
values of $N_\text{H,es}$.

As soon as the disk starts to open the line of sight to the neutron star
(indicated by the downward moving disk in the image ``Mid of Turn-On'' of
Fig.~\ref{fig:turnonmodel}), the parts visible of the corona increase and
consequently the observed flux in MC~I increases as well. In parallel, the
apparent absorbed flux increases, because the disk becomes more and more
transparent for photons scattered in the corona (beam~C).  Around the mid
point of the turn-on, the outer disk rim starts to become optically thin
for photons directly coming from the neutron star following beam~D (``Mid
of turn-on''). As a consequence the flux in component MC~II (the ratio $C$
in Fig.~\ref{fig:parameter}) rises more rapidly compared to the flux of
MC~I.  Since MC~II determines the curvature of the total observable
spectrum, $N_\text{H}$ apparently increases. Since the pulse profile due to
MC~II expected to be far less smeared, pulse profile fitting during this
episode will shown a decrease of the overall $N_\text{H,es}$. During the
further evolution the contribution of MC~II continues to increase until it
dominates the spectrum. This event takes place right after the turning
point in the track of $N_\text{H}$ shown in Fig.~\ref{fig:parameter}a.

Shortly before the anomalous dip, the apparent $N_\text{H}$ rapidly starts
to decrease. Both components MC~I and MC~II are now indistinguishable
(orbit~17 and later) because photoelectric absorption and electron
scattering become negligible.  Based on the interpretation of
\cite{cheng:95a}, this indicates that a larger amount of photons from the
inner parts of the accretion disk close to the neutron star reach the
observer. Finally, at the end of the turn-on, the neutron star is directly
visible when the main-on starts.

In summary, models as the one outlined above seem to be successful in
describing the overall features of the 35 day turn-on of Her~X-1. We
note, however, that our results in Sect.~\ref{sec:pulsesim} also show
that for very high columns, the separation into two components fails
due to low counting statistics during early phases of the turn-on
although the overall picture stays consistent.  The success of the
analysis presented here, however, is a first step towards
self-consistently modeling both, the spectral evolution and the timing
behavior of Her~X-1.  Further refinements of the method are still
possible, e.g., when taking the absolute phase shift of the observed
pulse profile into account and by using more realistic geometries.

\section{Acknowledgments}
We acknowledge partial funding from DLR grants 50~OX0002, 50~OR0302,
DFG grant Sta 173/31-1, and NATO grant PST.CL975254.

% Bibliographies
\bibliographystyle{aa}
\bibliography{mnemonic,her,diss,xte,accretion,general,radiation,conferences}

\appendix 
\section{Results of the spectral analysis}
\begin{sidewaystable*}
\renewcommand{\arraystretch}{0.7}
\footnotesize{
\caption{Results of the spectral fitting to the Her X-1 turn-on data with
  all parameters set free, except the parameters where no errors are given
  for and the power law index $\alpha$ which was kept fix at $\alpha=1.068$.
  \label{tab:results-spectral-analysis}}
\begin{center}
\begin{tabular}{cllllllllllllcr}
\hline\hline
Obs. &\multicolumn{1}{c}{$N_{\rm H}$} &\multicolumn{1}{c}{$C$}
&\multicolumn{1}{c}{$A_{\rm PL}$} & \multicolumn{1}{c}{$E_{\rm cut}$}
&\multicolumn{1}{c}{$E_{\rm fold}$} &\multicolumn{1}{c}{$E_{\rm Fe}$}
&\multicolumn{1}{c}{$\sigma_{\rm Fe}$} &\multicolumn{1}{c}{$A_{\rm Line}$}
&\multicolumn{1}{c}{$E_{\rm cyc}$} &\multicolumn{1}{c}{$\sigma_{\rm cyc}$}
&\multicolumn{1}{c}{$\tau_{\rm cyc}$} 
& \multicolumn{1}{c}{$\chi^2/\rm dof$} \\
 &\multicolumn{1}{c}{$10^{22}\rm{cm}^2$} &  &\multicolumn{1}{c}{$10^{-3}$} & \multicolumn{1}{c}{\kev} & \multicolumn{1}{c}{\kev} & \multicolumn{1}{c}{\kev} & \multicolumn{1}{c}{\kev} & \multicolumn{1}{c}{$10^{-4}$} &\multicolumn{1}{c}{\kev} & \multicolumn{1}{c}{\kev} &  &  \\
\hline
00
& $  44.73^{+  3.17}_{-  7.24}$ & $  2.14^{+ 0.14}_{- 0.13}$ & $   4.2^{+  0.2}_{-  0.2}$ & $ 18.16^{+ 3.46}_{- 2.21}$ & $  15.4^{+ 16.6}_{-  7.4}$ & $  6.69^{+ 0.11}_{- 0.14}$ & 0.77  
& $   4.6^{+  0.8}_{-  0.7}$ & 39.7  
& 5.10  
& $  0.50^{+ 2.58}_{- 0.50}$ & $ 77.3/ 103$ \\ \noalign{\vskip 2pt}
01
& $  47.34^{+  6.14}_{-  4.28}$ & $  2.19^{+ 0.15}_{- 0.12}$ & $   5.0^{+  0.2}_{-  0.2}$ & $ 18.24^{+ 2.21}_{- 1.57}$ & $   9.5^{+  5.6}_{-  3.3}$ & $  6.68^{+ 0.11}_{- 0.11}$ & 0.77  
& $   5.3^{+  0.7}_{-  0.7}$ & 39.7  
& 5.10  
& $  0.00^{+ 2.09}_{- 0.00}$ & $102.4/ 103$ \\ \noalign{\vskip 2pt}
02
& $  56.26^{+  4.13}_{-  3.65}$ & $  3.16^{+ 0.16}_{- 0.15}$ & $   6.1^{+  0.2}_{-  0.2}$ & $ 18.42^{+ 3.01}_{- 1.69}$ & $  20.0^{+ 10.3}_{-  7.5}$ & $  6.60^{+ 0.09}_{- 0.09}$ & 0.77  
& $   7.6^{+  0.8}_{-  0.8}$ & 39.7  
& 5.10  
& $  1.31^{+ 1.38}_{- 1.19}$ & $163.7/ 103$ \\ \noalign{\vskip 2pt}
03
& $  67.38^{+  1.07}_{-  7.67}$ & $  3.57^{+ 0.16}_{- 0.15}$ & $   8.1^{+  0.2}_{-  0.2}$ & $ 20.57^{+ 2.76}_{- 2.69}$ & $  14.3^{+  8.8}_{-  7.7}$ & $  6.57^{+ 0.07}_{- 0.07}$ & 0.77  
& $  11.4^{+  0.9}_{-  0.9}$ & 39.7  
& 5.10  
& $  1.61^{+ 1.71}_{- 1.61}$ & $134.5/ 103$ \\ \noalign{\vskip 2pt}
04
& $  77.35^{+  1.81}_{-  2.96}$ & $  4.49^{+ 0.14}_{- 0.15}$ & $  12.2^{+  0.2}_{-  0.2}$ & $ 19.98^{+ 1.45}_{- 1.63}$ & $  11.6^{+  3.9}_{-  2.8}$ & $  6.42^{+ 0.06}_{- 0.02}$ & 0.77  
& $  15.0^{+  1.1}_{-  1.0}$ & 39.7  
& 5.10  
& $  0.46^{+ 0.82}_{- 0.46}$ & $165.3/ 103$ \\ \noalign{\vskip 2pt}
05
& $  75.17^{+  1.62}_{-  2.39}$ & $  6.01^{+ 0.14}_{- 0.17}$ & $  15.2^{+  0.2}_{-  0.2}$ & $ 21.53^{+ 1.08}_{- 1.08}$ & $  13.6^{+  2.8}_{-  2.8}$ & $  6.54^{+ 0.07}_{- 0.08}$ & 0.77  
& $  16.5^{+  1.4}_{-  1.3}$ & 39.7  
& 5.10  
& $  0.91^{+ 0.56}_{- 0.56}$ & $198.9/ 103$ \\ \noalign{\vskip 2pt}
06
& $  61.37^{+  1.46}_{-  1.23}$ & $  7.15^{+ 0.13}_{- 0.15}$ & $  17.4^{+  0.3}_{-  0.3}$ & $ 21.05^{+ 0.92}_{- 0.98}$ & $  14.5^{+  2.4}_{-  2.2}$ & $  6.70^{+ 0.10}_{- 0.10}$ & $  1.00^{+ 0.13}_{- 0.02}$ & $  13.4^{+  1.8}_{-  1.8}$ & 39.7  
& 5.10  
& $  1.25^{+ 0.52}_{- 0.47}$ & $142.7/ 103$ \\ \noalign{\vskip 2pt}
07
& $  54.84^{+  1.35}_{-  1.26}$ & $  7.70^{+ 0.17}_{- 0.16}$ & $  17.7^{+  0.4}_{-  0.4}$ & $ 21.01^{+ 0.78}_{- 0.88}$ & $  12.0^{+  2.2}_{-  2.0}$ & $  6.71^{+ 0.09}_{- 0.12}$ & $  1.00^{+ 0.00}_{- 0.09}$ & $  17.7^{+  2.7}_{-  3.0}$ & 39.7  
& 5.10  
& $  0.41^{+ 0.51}_{- 0.41}$ & $122.2/ 102$ \\ \noalign{\vskip 2pt}
08
& $  65.75^{+  2.77}_{-  0.80}$ & $  9.94^{+ 0.31}_{- 0.20}$ & $  14.3^{+  0.3}_{-  0.3}$ & $ 21.07^{+ 0.86}_{- 0.93}$ & $  12.9^{+  2.2}_{-  1.9}$ & $  6.68^{+ 0.12}_{- 0.11}$ & $  0.98^{+ 0.13}_{- 0.02}$ & $  14.7^{+  2.3}_{-  1.7}$ & 39.7  
& 5.10  
& $  0.69^{+ 0.46}_{- 0.40}$ & $119.8/ 102$ \\ \noalign{\vskip 2pt}
09
& $  52.66^{+  1.27}_{-  1.39}$ & $  8.37^{+ 0.17}_{- 0.21}$ & $  17.6^{+  0.4}_{-  0.4}$ & $ 21.56^{+ 1.04}_{- 1.10}$ & $  11.3^{+  2.6}_{-  2.5}$ & $  6.67^{+ 0.13}_{- 0.15}$ & $  1.00^{+ 0.00}_{- 0.10}$ & $  17.1^{+  3.0}_{-  3.0}$ & 39.7  
& 5.10  
& $  0.77^{+ 0.67}_{- 0.68}$ & $135.1/ 102$ \\ \noalign{\vskip 2pt}
15
& $  25.36^{+  0.70}_{-  0.94}$ & $  5.75^{+ 0.13}_{- 0.13}$ & $  25.5^{+  0.8}_{-  1.4}$ & $ 21.24^{+ 0.39}_{- 0.41}$ & $  12.9^{+  1.0}_{-  1.0}$ & $  6.55^{+ 0.10}_{- 0.10}$ & $  1.00^{+ 0.00}_{- 0.60}$ & $  31.6^{+  4.2}_{-  4.6}$ & 39.7  
& 5.10  
& $  1.02^{+ 0.22}_{- 0.22}$ & $108.2/ 101$ \\ \noalign{\vskip 2pt}
16
& $  11.91^{+  2.30}_{-  2.32}$ & $  0.70^{+ 0.25}_{- 0.10}$ & $  98.3^{+  1.5}_{- 14.2}$ & $ 21.18^{+ 0.46}_{- 0.47}$ & $  12.6^{+  1.1}_{-  1.1}$ & $  6.55^{+ 0.10}_{- 0.11}$ & $  0.71^{+ 0.18}_{- 0.14}$ & $  39.2^{+  6.8}_{-  5.2}$ & 39.7  
& 5.10  
& $  0.80^{+ 0.25}_{- 0.24}$ & $ 93.1/ 101$ \\ \noalign{\vskip 2pt}
17
& $   4.74^{+  0.42}_{-  0.48}$ & 1.00    
& $  85.0^{+  0.4}_{-  0.4}$ & $ 21.24^{+ 0.46}_{- 0.48}$ & $  12.4^{+  1.1}_{-  1.1}$ & $  6.60^{+ 0.11}_{- 0.11}$ & $  0.72^{+ 0.18}_{- 0.15}$ & $  41.2^{+  6.6}_{-  5.5}$ & 39.7  
& 5.10  
& $  0.77^{+ 0.26}_{- 0.26}$ & $ 98.9/ 102$ \\ \noalign{\vskip 2pt}
18
& $   5.17^{+  0.45}_{-  0.46}$ & 1.00    
& $  85.3^{+  0.4}_{-  0.4}$ & $ 21.13^{+ 0.43}_{- 0.44}$ & $  12.0^{+  1.0}_{-  1.0}$ & $  6.63^{+ 0.11}_{- 0.12}$ & $  0.78^{+ 0.18}_{- 0.16}$ & $  41.2^{+  6.8}_{-  5.7}$ & 39.7  
& 5.10  
& $  0.76^{+ 0.25}_{- 0.25}$ & $ 84.4/ 102$ \\ \noalign{\vskip 2pt}
23
& $   6.49^{+  0.37}_{-  0.50}$ & 1.00    
& $ 100.5^{+  3.2}_{-  0.4}$ & $ 21.91^{+ 0.58}_{- 0.57}$ & $  10.8^{+  2.2}_{-  1.0}$ & $  6.53^{+ 0.11}_{- 0.13}$ & $  0.75^{+ 0.19}_{- 0.13}$ & $  50.4^{+  8.3}_{-  6.9}$ & $  40.6^{+  0.8}_{-  0.9}$ & 5.10  
& $  2.39^{+ 3.92}_{- 1.41}$ & $108.7/ 101$ \\ \noalign{\vskip 2pt}
24
& $   5.23^{+  0.48}_{-  0.38}$ & 1.00    
& $ 103.5^{+  0.7}_{-  0.5}$ & $ 20.77^{+ 0.67}_{- 0.71}$ & $  13.7^{+  1.6}_{-  1.3}$ & $  6.59^{+ 0.09}_{- 0.10}$ & $  0.70^{+ 0.18}_{- 0.15}$ & $  53.3^{+  8.7}_{-  7.1}$ & $  39.2^{+  1.1}_{-  1.0}$ & $  3.34^{+ 1.90}_{- 1.45}$ & $  1.48^{+ 0.96}_{- 0.45}$ & $122.7/ 100$ \\ \noalign{\vskip 2pt}
25
& $   4.30^{+  0.43}_{-  0.38}$ & 1.00    
& $ 104.8^{+  0.7}_{-  0.5}$ & $ 21.21^{+ 0.53}_{- 0.48}$ & $  13.4^{+  1.5}_{-  1.1}$ & $  6.62^{+ 0.09}_{- 0.10}$ & $  0.71^{+ 0.17}_{- 0.15}$ & $  52.8^{+  8.1}_{-  7.2}$ & $  39.7^{+  1.0}_{-  0.9}$ & $  5.37^{+ 2.13}_{- 1.50}$ & $  1.16^{+ 0.27}_{- 0.24}$ & $117.7/ 100$ \\ \noalign{\vskip 2pt}
26
& $   4.31^{+  0.40}_{-  0.38}$ & 1.00    
& $ 104.7^{+  0.6}_{-  0.4}$ & $ 22.06^{+ 0.38}_{- 0.34}$ & $  11.3^{+  1.1}_{-  0.9}$ & $  6.64^{+ 0.08}_{- 0.10}$ & $  0.78^{+ 0.17}_{- 0.13}$ & $  58.8^{+  8.6}_{-  6.7}$ & $  40.0^{+  1.2}_{-  1.2}$ & $  5.18^{+ 2.85}_{- 2.32}$ & $  0.71^{+ 0.20}_{- 0.17}$ & $106.6/ 100$ \\ \noalign{\vskip 2pt}
27
& $   4.58^{+  0.41}_{-  0.39}$ & 1.00    
& $ 110.9^{+  0.7}_{-  0.5}$ & $ 21.01^{+ 0.35}_{- 0.36}$ & $  12.8^{+  1.3}_{-  0.9}$ & $  6.68^{+ 0.09}_{- 0.10}$ & $  0.75^{+ 0.17}_{- 0.14}$ & $  57.2^{+  8.4}_{-  7.0}$ & $  40.4^{+  0.9}_{-  0.9}$ & $  5.16^{+ 2.33}_{- 1.69}$ & $  0.95^{+ 0.19}_{- 0.17}$ & $110.9/ 100$ \\ \noalign{\vskip 2pt}
28
& $   4.36^{+  0.36}_{-  0.35}$ & 1.00    
& $ 110.5^{+  3.9}_{-  0.7}$ & $ 21.19^{+ 0.29}_{- 0.29}$ & $  12.2^{+  0.8}_{-  0.6}$ & $  6.65^{+ 0.08}_{- 0.09}$ & $  0.79^{+ 0.17}_{- 0.15}$ & $  61.9^{+  7.9}_{-  7.5}$ & $  39.1^{+  0.8}_{-  0.8}$ & $  4.70^{+ 1.81}_{- 1.33}$ & $  0.83^{+ 0.16}_{- 0.15}$ & $114.6/ 100$ \\ \noalign{\vskip 2pt}
29
& $   4.24^{+  0.36}_{-  0.31}$ & 1.00    
& $ 116.1^{+  0.7}_{-  0.6}$ & $ 21.40^{+ 0.33}_{- 0.31}$ & $  12.1^{+  1.0}_{-  0.6}$ & $  6.68^{+ 0.08}_{- 0.10}$ & $  0.83^{+ 0.18}_{- 0.16}$ & $  65.8^{+  7.8}_{-  8.2}$ & $  39.7^{+  1.0}_{-  1.0}$ & $  5.59^{+ 3.31}_{- 1.55}$ & $  0.71^{+ 0.14}_{- 0.13}$ & $114.5/ 100$ \\ \noalign{\vskip 2pt}
30
& $   4.55^{+  0.34}_{-  0.41}$ & 1.00    
& $ 118.3^{+  0.7}_{-  0.5}$ & $ 21.14^{+ 0.39}_{- 0.37}$ & $  12.7^{+  1.2}_{-  0.9}$ & $  6.67^{+ 0.08}_{- 0.08}$ & $  0.77^{+ 0.03}_{- 0.15}$ & $  64.3^{+  6.4}_{-  6.9}$ & $  39.5^{+  0.7}_{-  0.7}$ & $  5.25^{+ 1.92}_{- 1.48}$ & $  1.05^{+ 0.20}_{- 0.19}$ & $119.7/ 100$ \\ \noalign{\vskip 2pt}
31
& $   3.48^{+  0.37}_{-  0.37}$ & 1.00    
& $ 116.6^{+  0.5}_{-  0.4}$ & $ 21.37^{+ 0.44}_{- 0.46}$ & $  11.6^{+  0.7}_{-  0.7}$ & $  6.59^{+ 0.07}_{- 0.08}$ & $  0.77^{+ 0.03}_{- 0.13}$ & $  71.2^{+  7.1}_{-  7.5}$ & $  40.5^{+  0.7}_{-  0.7}$ & $  1.00^{+ 6.25}_{- 0.00}$ & $  1.97^{+ 0.99}_{- 1.39}$ & $ 95.8/ 100$ \\ \noalign{\vskip 2pt}
\hline
\end{tabular}
\end{center}
{\small 
$N_{\rm H}$: Hydrogen column density, 
$C$: Relative ratio of absorbed and scattered radiation to the unaffected radiation,
$A_{\rm PL}$: Power law normalization,
($\text{photons}\,\text{cm}^{-2}\,\text{s}^{-1}\,\text{keV}^{-1}$ at $1\,\text{\kev}$), 
$A_{\rm Line}$: Line normalization
($\text{photons}\,\text{cm}^{-2}\,\text{s}^{-1}$ in the line), 
$E_{\rm cut}$: Cut-off energy (\kev), 
$E_{\rm fold}$: Folding energy (\kev), 
$E_{\rm Fe}$: Gaussian line energy (\kev), 
$\sigma_{\rm Fe}$: Width of the Gaussian line (\kev), 
$A_{\rm Line}$: Line normalization, 
$E_{\rm cyc}$: Cyclotron line energy (\kev), 
$\sigma_{\rm cyc}$: Width of the cyclotron line (\kev), 
$\tau_{\rm cyc}$: Depth of the cyclotron line.
Uncertainties are at $90\,\%$ confidence level for one interesting
parameter ($\Delta \chi^2 = 2.71$).
}
}
\end{sidewaystable*}

%%% Local Variables: 
%%% mode: latex
%%% TeX-master: "~/publ/diss/tables/../diss"
%%% End: 
 
\begin{table*}
\renewcommand{\arraystretch}{1.0}
\caption{Results of the spectral fitting to the Her X-1 turn-on data for the following parameters: $N_{\rm H}$, $C$, $A_{\rm PL}$, and $A_{Line}$.\label{tab:her_short_fits}}
\begin{center}
\begin{tabular}{cllllcr}\hline\hline
Obs. & \multicolumn{1}{c}{$N_{\rm H}$} & \multicolumn{1}{c}{$C$} &\multicolumn{1}{c}{$A_{\rm PL}$} & \multicolumn{1}{c}{$A_{\rm Line}$} & \multicolumn{1}{c}{$\chi^2/\rm dof$} \\
 & \multicolumn{1}{c}{$10^{22}\rm{cm}^2$} & & \multicolumn{1}{c}{$10^{-3}$} &\multicolumn{1}{c}{$10^{-4}$} & & \\
\hline
00
& $ 35.98^{+ 7.60}_{- 1.93}$ & $   2.10^{+  0.16}_{-  0.15}$ & $ ~  4.2^{+  0.2}_{-  0.2}$ & $ ~  3.8^{+  0.6}_{-  0.6}$ & $ 83.9/ 108$ \\ \noalign{\vskip 2pt}
01
& $ 42.20^{+ 6.44}_{- 2.63}$ & $   2.13^{+  0.14}_{-  0.13}$ & $ ~  4.9^{+  0.2}_{-  0.2}$ & $ ~  4.1^{+  0.6}_{-  0.6}$ & $112.4/ 108$ \\ \noalign{\vskip 2pt}
02
& $ 62.80^{+ 4.71}_{- 3.65}$ & $   3.59^{+  0.23}_{-  0.21}$ & $ ~  6.1^{+  0.2}_{-  0.2}$ & $  10.7^{+  1.1}_{-  1.1}$ & $162.3/ 108$ \\ \noalign{\vskip 2pt}
03
& $ 67.61^{+ 4.17}_{- 2.60}$ & $   3.97^{+  0.20}_{-  0.19}$ & $ ~  8.0^{+  0.2}_{-  0.2}$ & $  14.3^{+  1.1}_{-  1.1}$ & $131.1/ 108$ \\ \noalign{\vskip 2pt}
04
& $ 83.92^{+ 0.33}_{- 4.83}$ & $   5.11^{+  0.21}_{-  0.19}$ & $  11.8^{+  0.2}_{-  0.2}$ & $  20.4^{+  1.4}_{-  1.4}$ & $151.1/ 108$ \\ \noalign{\vskip 2pt}
05
& $ 82.43^{+ 1.91}_{- 2.03}$ & $   7.61^{+  0.28}_{-  0.28}$ & $  13.9^{+  0.3}_{-  0.3}$ & $  35.3^{+  2.5}_{-  2.5}$ & $115.5/ 108$ \\ \noalign{\vskip 2pt}
06
& $ 66.27^{+ 1.65}_{- 1.65}$ & $   8.19^{+  0.26}_{-  0.25}$ & $  16.6^{+  0.3}_{-  0.3}$ & $  31.0^{+  3.6}_{-  3.7}$ & $130.4/ 108$ \\ \noalign{\vskip 2pt}
07
& $ 55.18^{+ 1.14}_{- 1.57}$ & $   7.80^{+  0.18}_{-  0.17}$ & $  17.7^{+  0.3}_{-  0.4}$ & $  19.0^{+  2.8}_{-  2.8}$ & $138.9/ 108$ \\ \noalign{\vskip 2pt}
08
& $ 67.47^{+ 0.74}_{- 2.29}$ & $  10.10^{+  0.25}_{-  0.26}$ & $  14.2^{+  0.3}_{-  0.3}$ & $  16.0^{+  2.5}_{-  2.5}$ & $135.3/ 108$ \\ \noalign{\vskip 2pt}
09
& $ 53.46^{+ 1.84}_{- 1.09}$ & $   8.64^{+  0.22}_{-  0.21}$ & $  17.5^{+  0.1}_{-  0.4}$ & $  22.2^{+  2.3}_{-  3.8}$ & $154.4/ 108$ \\ \noalign{\vskip 2pt}
15
& $ 24.02^{+ 0.81}_{- 0.52}$ & $   6.00^{+  0.30}_{-  0.24}$ & $  24.4^{+  1.0}_{-  1.1}$ & $  23.0^{+  2.8}_{-  2.8}$ & $145.7/ 107$ \\ \noalign{\vskip 2pt}
16
& $  8.13^{+ 4.50}_{- 0.90}$ & $   0.77^{+  0.50}_{-  0.10}$ & $  94.0^{+  1.0}_{-  2.2}$ & $  34.0^{+  3.7}_{-  3.4}$ & $128.7/ 107$ \\ \noalign{\vskip 2pt}
17
& $  4.84^{+ 0.37}_{- 0.44}$ & $   1.00^{+  0.00}_{-  0.00}$ & $  85.5^{+  0.4}_{-  0.4}$ & $  39.6^{+  3.9}_{-  3.9}$ & $135.6/ 108$ \\ \noalign{\vskip 2pt}
18
& $  5.34^{+ 0.43}_{- 0.41}$ & $   1.00^{+  0.00}_{-  0.00}$ & $  85.9^{+  0.4}_{-  0.4}$ & $  38.5^{+  3.9}_{-  3.9}$ & $105.6/ 108$ \\ \noalign{\vskip 2pt}
23
& $  6.72^{+ 0.54}_{- 0.35}$ & $   1.00^{+  0.00}_{-  0.00}$ & $ 101.3^{+  0.6}_{-  0.5}$ & $  48.7^{+  5.1}_{-  5.0}$ & $121.9/ 108$ \\ \noalign{\vskip 2pt}
24
& $  5.30^{+ 0.43}_{- 0.31}$ & $   1.00^{+  0.00}_{-  0.00}$ & $ 104.2^{+  0.6}_{-  0.5}$ & $  46.6^{+  4.9}_{-  4.8}$ & $144.1/ 108$ \\ \noalign{\vskip 2pt}
25
& $  4.41^{+ 0.40}_{- 0.38}$ & $   1.00^{+  0.00}_{-  0.00}$ & $ 105.2^{+  4.1}_{-  0.4}$ & $  51.2^{+  4.9}_{-  4.9}$ & $137.4/ 108$ \\ \noalign{\vskip 2pt}
26
& $  4.55^{+ 0.35}_{- 0.42}$ & $   1.00^{+  0.00}_{-  0.00}$ & $ 105.4^{+  0.5}_{-  0.4}$ & $  55.2^{+  4.7}_{-  4.7}$ & $129.3/ 108$ \\ \noalign{\vskip 2pt}
27
& $  4.72^{+ 0.35}_{- 0.42}$ & $   1.00^{+  0.00}_{-  0.00}$ & $ 111.4^{+  0.5}_{-  0.4}$ & $  54.8^{+  4.8}_{-  4.8}$ & $134.8/ 108$ \\ \noalign{\vskip 2pt}
28
& $  4.47^{+ 0.37}_{- 0.39}$ & $   1.00^{+  0.00}_{-  0.00}$ & $ 111.2^{+  0.5}_{-  0.5}$ & $  58.0^{+  5.0}_{-  5.0}$ & $140.9/ 108$ \\ \noalign{\vskip 2pt}
29
& $  4.33^{+ 0.36}_{- 0.40}$ & $   1.00^{+  0.00}_{-  0.00}$ & $ 116.6^{+  0.5}_{-  0.5}$ & $  63.8^{+  5.4}_{-  5.4}$ & $173.8/ 108$ \\ \noalign{\vskip 2pt}
30
& $  4.75^{+ 0.35}_{- 0.43}$ & $   1.00^{+  0.00}_{-  0.00}$ & $ 156.7^{+ 17.1}_{- 14.1}$ & $  65.1^{+  6.4}_{-  5.8}$ & $150.2/ 108$ \\ \noalign{\vskip 2pt}
31
& $  3.70^{+ 0.40}_{- 0.38}$ & $   1.00^{+  0.00}_{-  0.00}$ & $ 117.4^{+  0.6}_{-  0.5}$ & $  70.7^{+  5.8}_{-  5.8}$ & $102.9/ 108$ \\ \noalign{\vskip 2pt}
\hline
\end{tabular}
\end{center}
{\small 
$N_{\rm H}$: Hydrogen column density,
C: Relative ratio of absorbed and scattered radiation to the unaffected radiation,
$A_{\rm PL}$: Power law normalization
(photons\,cm$^{-2}$\,s$^{-1}$\,keV$^{-1}$ at 1\,keV),
$A_{\rm Line}$: Line normalization
(photons\,cm$^{-2}$\,s$^{-1}$ in the line), the following parameters were fixed:
the Gaussian emission line was fixed at 6.45\,keV with a width $\sigma$ of 0.45\,keV, the cut off 
energy $E_{\rm cut}$ at 21.5\,\kev, the folding energy $E_{\rm fold}$ at 14.1\,\kev, the
power law index $\alpha$ at 1.068, and the cyclotron energy
$E_{\rm cyc}$ at 39.4\,\kev with a width of $5.1\,\kev$.
Uncertainties are at 90\% confidence level for one interesting
parameter ($\Delta \chi^2 = 2.71$).
}
\end{table*}

%%% Local Variables: 
%%% mode: latex
%%% TeX-master: "~/publ/diss/diss"
%%% End: 

\end{document}